
\catcode`\@=11


\message{Loading jyTeX fonts...}



\font\vptrm=cmr5 \font\vptmit=cmmi5 \font\vptsy=cmsy5
\font\vptbf=cmbx5

\skewchar\vptmit='177 \skewchar\vptsy='60 \fontdimen16
\vptsy=\the\fontdimen17 \vptsy

\def\vpt{\ifmmode\err@badsizechange\else
     \@mathfontinit
     \textfont0=\vptrm  \scriptfont0=\vptrm  \scriptscriptfont0=\vptrm
     \textfont1=\vptmit \scriptfont1=\vptmit \scriptscriptfont1=\vptmit
     \textfont2=\vptsy  \scriptfont2=\vptsy  \scriptscriptfont2=\vptsy
     \textfont3=\xptex  \scriptfont3=\xptex  \scriptscriptfont3=\xptex
     \textfont\bffam=\vptbf
     \scriptfont\bffam=\vptbf
     \scriptscriptfont\bffam=\vptbf
     \@fontstyleinit
     \def\rm{\vptrm\fam=\z@}%
     \def\bf{\vptbf\fam=\bffam}%
     \def\oldstyle{\vptmit\fam=\@ne}%
     \rm\fi}


\font\viptrm=cmr6 \font\viptmit=cmmi6 \font\viptsy=cmsy6
\font\viptbf=cmbx6

\skewchar\viptmit='177 \skewchar\viptsy='60 \fontdimen16
\viptsy=\the\fontdimen17 \viptsy

\def\vipt{\ifmmode\err@badsizechange\else
     \@mathfontinit
     \textfont0=\viptrm  \scriptfont0=\vptrm  \scriptscriptfont0=\vptrm
     \textfont1=\viptmit \scriptfont1=\vptmit \scriptscriptfont1=\vptmit
     \textfont2=\viptsy  \scriptfont2=\vptsy  \scriptscriptfont2=\vptsy
     \textfont3=\xptex   \scriptfont3=\xptex  \scriptscriptfont3=\xptex
     \textfont\bffam=\viptbf
     \scriptfont\bffam=\vptbf
     \scriptscriptfont\bffam=\vptbf
     \@fontstyleinit
     \def\rm{\viptrm\fam=\z@}%
     \def\bf{\viptbf\fam=\bffam}%
     \def\oldstyle{\viptmit\fam=\@ne}%
     \rm\fi}


\font\viiptrm=cmr7 \font\viiptmit=cmmi7 \font\viiptsy=cmsy7
\font\viiptit=cmti7 \font\viiptbf=cmbx7

\skewchar\viiptmit='177 \skewchar\viiptsy='60 \fontdimen16
\viiptsy=\the\fontdimen17 \viiptsy

\def\viipt{\ifmmode\err@badsizechange\else
     \@mathfontinit
     \textfont0=\viiptrm  \scriptfont0=\vptrm  \scriptscriptfont0=\vptrm
     \textfont1=\viiptmit \scriptfont1=\vptmit \scriptscriptfont1=\vptmit
     \textfont2=\viiptsy  \scriptfont2=\vptsy  \scriptscriptfont2=\vptsy
     \textfont3=\xptex    \scriptfont3=\xptex  \scriptscriptfont3=\xptex
     \textfont\itfam=\viiptit
     \scriptfont\itfam=\viiptit
     \scriptscriptfont\itfam=\viiptit
     \textfont\bffam=\viiptbf
     \scriptfont\bffam=\vptbf
     \scriptscriptfont\bffam=\vptbf
     \@fontstyleinit
     \def\rm{\viiptrm\fam=\z@}%
     \def\it{\viiptit\fam=\itfam}%
     \def\bf{\viiptbf\fam=\bffam}%
     \def\oldstyle{\viiptmit\fam=\@ne}%
     \rm\fi}


\font\viiiptrm=cmr8 \font\viiiptmit=cmmi8 \font\viiiptsy=cmsy8
\font\viiiptit=cmti8
\font\viiiptbf=cmbx8

\skewchar\viiiptmit='177 \skewchar\viiiptsy='60 \fontdimen16
\viiiptsy=\the\fontdimen17 \viiiptsy

\def\viiipt{\ifmmode\err@badsizechange\else
     \@mathfontinit
     \textfont0=\viiiptrm  \scriptfont0=\viptrm  \scriptscriptfont0=\vptrm
     \textfont1=\viiiptmit \scriptfont1=\viptmit \scriptscriptfont1=\vptmit
     \textfont2=\viiiptsy  \scriptfont2=\viptsy  \scriptscriptfont2=\vptsy
     \textfont3=\xptex     \scriptfont3=\xptex   \scriptscriptfont3=\xptex
     \textfont\itfam=\viiiptit
     \scriptfont\itfam=\viiptit
     \scriptscriptfont\itfam=\viiptit
     \textfont\bffam=\viiiptbf
     \scriptfont\bffam=\viptbf
     \scriptscriptfont\bffam=\vptbf
     \@fontstyleinit
     \def\rm{\viiiptrm\fam=\z@}%
     \def\it{\viiiptit\fam=\itfam}%
     \def\bf{\viiiptbf\fam=\bffam}%
     \def\oldstyle{\viiiptmit\fam=\@ne}%
     \rm\fi}


\def\getixpt{%
     \font\ixptrm=cmr9
     \font\ixptmit=cmmi9
     \font\ixptsy=cmsy9
     \font\ixptit=cmti9
     \font\ixptbf=cmbx9
     \skewchar\ixptmit='177 \skewchar\ixptsy='60
     \fontdimen16 \ixptsy=\the\fontdimen17 \ixptsy}

\def\ixpt{\ifmmode\err@badsizechange\else
     \@mathfontinit
     \textfont0=\ixptrm  \scriptfont0=\viiptrm  \scriptscriptfont0=\vptrm
     \textfont1=\ixptmit \scriptfont1=\viiptmit \scriptscriptfont1=\vptmit
     \textfont2=\ixptsy  \scriptfont2=\viiptsy  \scriptscriptfont2=\vptsy
     \textfont3=\xptex   \scriptfont3=\xptex    \scriptscriptfont3=\xptex
     \textfont\itfam=\ixptit
     \scriptfont\itfam=\viiptit
     \scriptscriptfont\itfam=\viiptit
     \textfont\bffam=\ixptbf
     \scriptfont\bffam=\viiptbf
     \scriptscriptfont\bffam=\vptbf
     \@fontstyleinit
     \def\rm{\ixptrm\fam=\z@}%
     \def\it{\ixptit\fam=\itfam}%
     \def\bf{\ixptbf\fam=\bffam}%
     \def\oldstyle{\ixptmit\fam=\@ne}%
     \rm\fi}


\font\xptrm=cmr10 \font\xptmit=cmmi10 \font\xptsy=cmsy10
\font\xptex=cmex10 \font\xptit=cmti10 \font\xptsl=cmsl10
\font\xptbf=cmbx10 \font\xpttt=cmtt10 \font\xptss=cmss10
\font\xptsc=cmcsc10 \font\xptbfs=cmb10 \font\xptbmit=cmmib10

\skewchar\xptmit='177 \skewchar\xptbmit='177 \skewchar\xptsy='60
\fontdimen16 \xptsy=\the\fontdimen17 \xptsy

\def\xpt{\ifmmode\err@badsizechange\else
     \@mathfontinit
     \textfont0=\xptrm  \scriptfont0=\viiptrm  \scriptscriptfont0=\vptrm
     \textfont1=\xptmit \scriptfont1=\viiptmit \scriptscriptfont1=\vptmit
     \textfont2=\xptsy  \scriptfont2=\viiptsy  \scriptscriptfont2=\vptsy
     \textfont3=\xptex  \scriptfont3=\xptex    \scriptscriptfont3=\xptex
     \textfont\itfam=\xptit
     \scriptfont\itfam=\viiptit
     \scriptscriptfont\itfam=\viiptit
     \textfont\bffam=\xptbf
     \scriptfont\bffam=\viiptbf
     \scriptscriptfont\bffam=\vptbf
     \textfont\bfsfam=\xptbfs
     \scriptfont\bfsfam=\viiptbf
     \scriptscriptfont\bfsfam=\vptbf
     \textfont\bmitfam=\xptbmit
     \scriptfont\bmitfam=\viiptmit
     \scriptscriptfont\bmitfam=\vptmit
     \@fontstyleinit
     \def\rm{\xptrm\fam=\z@}%
     \def\it{\xptit\fam=\itfam}%
     \def\sl{\xptsl}%
     \def\bf{\xptbf\fam=\bffam}%
     \def\tt{\xpttt}%
     \def\ss{\xptss}%
     \def\sc{\xptsc}%
     \def\bfs{\xptbfs\fam=\bfsfam}%
     \def\bmit{\fam=\bmitfam}%
     \def\oldstyle{\xptmit\fam=\@ne}%
     \rm\fi}


\def\getxipt{%
     \font\xiptrm=cmr10  scaled\magstephalf
     \font\xiptmit=cmmi10 scaled\magstephalf
     \font\xiptsy=cmsy10 scaled\magstephalf
     \font\xiptex=cmex10 scaled\magstephalf
     \font\xiptit=cmti10 scaled\magstephalf
     \font\xiptsl=cmsl10 scaled\magstephalf
     \font\xiptbf=cmbx10 scaled\magstephalf
     \font\xipttt=cmtt10 scaled\magstephalf
     \font\xiptss=cmss10 scaled\magstephalf
     \skewchar\xiptmit='177 \skewchar\xiptsy='60
     \fontdimen16 \xiptsy=\the\fontdimen17 \xiptsy}

\def\xipt{\ifmmode\err@badsizechange\else
     \@mathfontinit
     \textfont0=\xiptrm  \scriptfont0=\viiiptrm  \scriptscriptfont0=\viptrm
     \textfont1=\xiptmit \scriptfont1=\viiiptmit \scriptscriptfont1=\viptmit
     \textfont2=\xiptsy  \scriptfont2=\viiiptsy  \scriptscriptfont2=\viptsy
     \textfont3=\xiptex  \scriptfont3=\xptex     \scriptscriptfont3=\xptex
     \textfont\itfam=\xiptit
     \scriptfont\itfam=\viiiptit
     \scriptscriptfont\itfam=\viiptit
     \textfont\bffam=\xiptbf
     \scriptfont\bffam=\viiiptbf
     \scriptscriptfont\bffam=\viptbf
     \@fontstyleinit
     \def\rm{\xiptrm\fam=\z@}%
     \def\it{\xiptit\fam=\itfam}%
     \def\sl{\xiptsl}%
     \def\bf{\xiptbf\fam=\bffam}%
     \def\tt{\xipttt}%
     \def\ss{\xiptss}%
     \def\oldstyle{\xiptmit\fam=\@ne}%
     \rm\fi}


\font\xiiptrm=cmr12 \font\xiiptmit=cmmi12 \font\xiiptsy=cmsy10
scaled\magstep1 \font\xiiptex=cmex10  scaled\magstep1
\font\xiiptit=cmti12 \font\xiiptsl=cmsl12 \font\xiiptbf=cmbx12
\font\xiiptss=cmss12 \font\xiiptsc=cmcsc10 scaled\magstep1
\font\xiiptbfs=cmb10  scaled\magstep1 \font\xiiptbmit=cmmib10
scaled\magstep1

\skewchar\xiiptmit='177 \skewchar\xiiptbmit='177
\skewchar\xiiptsy='60 \fontdimen16 \xiiptsy=\the\fontdimen17 \xiiptsy

\def\xiipt{\ifmmode\err@badsizechange\else
     \@mathfontinit
     \textfont0=\xiiptrm  \scriptfont0=\viiiptrm  \scriptscriptfont0=\viptrm
     \textfont1=\xiiptmit \scriptfont1=\viiiptmit \scriptscriptfont1=\viptmit
     \textfont2=\xiiptsy  \scriptfont2=\viiiptsy  \scriptscriptfont2=\viptsy
     \textfont3=\xiiptex  \scriptfont3=\xptex     \scriptscriptfont3=\xptex
     \textfont\itfam=\xiiptit
     \scriptfont\itfam=\viiiptit
     \scriptscriptfont\itfam=\viiptit
     \textfont\bffam=\xiiptbf
     \scriptfont\bffam=\viiiptbf
     \scriptscriptfont\bffam=\viptbf
     \textfont\bfsfam=\xiiptbfs
     \scriptfont\bfsfam=\viiiptbf
     \scriptscriptfont\bfsfam=\viptbf
     \textfont\bmitfam=\xiiptbmit
     \scriptfont\bmitfam=\viiiptmit
     \scriptscriptfont\bmitfam=\viptmit
     \@fontstyleinit
     \def\rm{\xiiptrm\fam=\z@}%
     \def\it{\xiiptit\fam=\itfam}%
     \def\sl{\xiiptsl}%
     \def\bf{\xiiptbf\fam=\bffam}%
     \def\tt{\xiipttt}%
     \def\ss{\xiiptss}%
     \def\sc{\xiiptsc}%
     \def\bfs{\xiiptbfs\fam=\bfsfam}%
     \def\bmit{\fam=\bmitfam}%
     \def\oldstyle{\xiiptmit\fam=\@ne}%
     \rm\fi}


\def\getxiiipt{%
     \font\xiiiptrm=cmr12  scaled\magstephalf
     \font\xiiiptmit=cmmi12 scaled\magstephalf
     \font\xiiiptsy=cmsy9  scaled\magstep2
     \font\xiiiptit=cmti12 scaled\magstephalf
     \font\xiiiptsl=cmsl12 scaled\magstephalf
     \font\xiiiptbf=cmbx12 scaled\magstephalf
     \font\xiiipttt=cmtt12 scaled\magstephalf
     \font\xiiiptss=cmss12 scaled\magstephalf
     \skewchar\xiiiptmit='177 \skewchar\xiiiptsy='60
     \fontdimen16 \xiiiptsy=\the\fontdimen17 \xiiiptsy}

\def\xiiipt{\ifmmode\err@badsizechange\else
     \@mathfontinit
     \textfont0=\xiiiptrm  \scriptfont0=\xptrm  \scriptscriptfont0=\viiptrm
     \textfont1=\xiiiptmit \scriptfont1=\xptmit \scriptscriptfont1=\viiptmit
     \textfont2=\xiiiptsy  \scriptfont2=\xptsy  \scriptscriptfont2=\viiptsy
     \textfont3=\xivptex   \scriptfont3=\xptex  \scriptscriptfont3=\xptex
     \textfont\itfam=\xiiiptit
     \scriptfont\itfam=\xptit
     \scriptscriptfont\itfam=\viiptit
     \textfont\bffam=\xiiiptbf
     \scriptfont\bffam=\xptbf
     \scriptscriptfont\bffam=\viiptbf
     \@fontstyleinit
     \def\rm{\xiiiptrm\fam=\z@}%
     \def\it{\xiiiptit\fam=\itfam}%
     \def\sl{\xiiiptsl}%
     \def\bf{\xiiiptbf\fam=\bffam}%
     \def\tt{\xiiipttt}%
     \def\ss{\xiiiptss}%
     \def\oldstyle{\xiiiptmit\fam=\@ne}%
     \rm\fi}


\font\xivptrm=cmr12   scaled\magstep1 \font\xivptmit=cmmi12
scaled\magstep1 \font\xivptsy=cmsy10  scaled\magstep2
\font\xivptex=cmex10  scaled\magstep2 \font\xivptit=cmti12
scaled\magstep1 \font\xivptsl=cmsl12  scaled\magstep1
\font\xivptbf=cmbx12  scaled\magstep1
\font\xivptss=cmss12  scaled\magstep1 \font\xivptsc=cmcsc10
scaled\magstep2 \font\xivptbfs=cmb10  scaled\magstep2
\font\xivptbmit=cmmib10 scaled\magstep2

\skewchar\xivptmit='177 \skewchar\xivptbmit='177
\skewchar\xivptsy='60 \fontdimen16 \xivptsy=\the\fontdimen17 \xivptsy

\def\xivpt{\ifmmode\err@badsizechange\else
     \@mathfontinit
     \textfont0=\xivptrm  \scriptfont0=\xptrm  \scriptscriptfont0=\viiptrm
     \textfont1=\xivptmit \scriptfont1=\xptmit \scriptscriptfont1=\viiptmit
     \textfont2=\xivptsy  \scriptfont2=\xptsy  \scriptscriptfont2=\viiptsy
     \textfont3=\xivptex  \scriptfont3=\xptex  \scriptscriptfont3=\xptex
     \textfont\itfam=\xivptit
     \scriptfont\itfam=\xptit
     \scriptscriptfont\itfam=\viiptit
     \textfont\bffam=\xivptbf
     \scriptfont\bffam=\xptbf
     \scriptscriptfont\bffam=\viiptbf
     \textfont\bfsfam=\xivptbfs
     \scriptfont\bfsfam=\xptbfs
     \scriptscriptfont\bfsfam=\viiptbf
     \textfont\bmitfam=\xivptbmit
     \scriptfont\bmitfam=\xptbmit
     \scriptscriptfont\bmitfam=\viiptmit
     \@fontstyleinit
     \def\rm{\xivptrm\fam=\z@}%
     \def\it{\xivptit\fam=\itfam}%
     \def\sl{\xivptsl}%
     \def\bf{\xivptbf\fam=\bffam}%
     \def\tt{\xivpttt}%
     \def\ss{\xivptss}%
     \def\sc{\xivptsc}%
     \def\bfs{\xivptbfs\fam=\bfsfam}%
     \def\bmit{\fam=\bmitfam}%
     \def\oldstyle{\xivptmit\fam=\@ne}%
     \rm\fi}


\font\xviiptrm=cmr17 \font\xviiptmit=cmmi12 scaled\magstep2
\font\xviiptsy=cmsy10 scaled\magstep3 \font\xviiptex=cmex10
scaled\magstep3 \font\xviiptit=cmti12 scaled\magstep2
\font\xviiptbf=cmbx12 scaled\magstep2 \font\xviiptbfs=cmb10
scaled\magstep3

\skewchar\xviiptmit='177 \skewchar\xviiptsy='60 \fontdimen16
\xviiptsy=\the\fontdimen17 \xviiptsy

\def\xviipt{\ifmmode\err@badsizechange\else
     \@mathfontinit
     \textfont0=\xviiptrm  \scriptfont0=\xiiptrm  \scriptscriptfont0=\viiiptrm
     \textfont1=\xviiptmit \scriptfont1=\xiiptmit \scriptscriptfont1=\viiiptmit
     \textfont2=\xviiptsy  \scriptfont2=\xiiptsy  \scriptscriptfont2=\viiiptsy
     \textfont3=\xviiptex  \scriptfont3=\xiiptex  \scriptscriptfont3=\xptex
     \textfont\itfam=\xviiptit
     \scriptfont\itfam=\xiiptit
     \scriptscriptfont\itfam=\viiiptit
     \textfont\bffam=\xviiptbf
     \scriptfont\bffam=\xiiptbf
     \scriptscriptfont\bffam=\viiiptbf
     \textfont\bfsfam=\xviiptbfs
     \scriptfont\bfsfam=\xiiptbfs
     \scriptscriptfont\bfsfam=\viiiptbf
     \@fontstyleinit
     \def\rm{\xviiptrm\fam=\z@}%
     \def\it{\xviiptit\fam=\itfam}%
     \def\bf{\xviiptbf\fam=\bffam}%
     \def\bfs{\xviiptbfs\fam=\bfsfam}%
     \def\oldstyle{\xviiptmit\fam=\@ne}%
     \rm\fi}


\font\xxiptrm=cmr17  scaled\magstep1


\def\xxipt{\ifmmode\err@badsizechange\else
     \@mathfontinit
     \@fontstyleinit
     \def\rm{\xxiptrm\fam=\z@}%
     \rm\fi}


\font\xxvptrm=cmr17  scaled\magstep2


\def\xxvpt{\ifmmode\err@badsizechange\else
     \@mathfontinit
     \@fontstyleinit
     \def\rm{\xxvptrm\fam=\z@}%
     \rm\fi}




\message{Loading jyTeX macros...}

\message{modifications to plain.tex,}


\def\newcount{\alloc@0\count\countdef\insc@unt}
\def\newdimen{\alloc@1\dimen\dimendef\insc@unt}
\def\newskip{\alloc@2\skip\skipdef\insc@unt}
\def\newmuskip{\alloc@3\muskip\muskipdef\@cclvi}
\def\newbox{\alloc@4\box\chardef\insc@unt}
\def\newtoks{\alloc@5\toks\toksdef\@cclvi}
\def\newhelp#1#2{\newtoks#1\global#1\expandafter{\csname#2\endcsname}}
\def\newread{\alloc@6\read\chardef\sixt@@n}
\def\newwrite{\alloc@7\write\chardef\sixt@@n}
\def\newfam{\alloc@8\fam\chardef\sixt@@n}
\def\newinsert#1{\global\advance\insc@unt by\m@ne
     \ch@ck0\insc@unt\count
     \ch@ck1\insc@unt\dimen
     \ch@ck2\insc@unt\skip
     \ch@ck4\insc@unt\box
     \allocationnumber=\insc@unt
     \global\chardef#1=\allocationnumber
     \wlog{\string#1=\string\insert\the\allocationnumber}}
\def\newif#1{\count@\escapechar \escapechar\m@ne
     \expandafter\expandafter\expandafter
          \xdef\@if#1{true}{\let\noexpand#1=\noexpand\iftrue}%
     \expandafter\expandafter\expandafter
          \xdef\@if#1{false}{\let\noexpand#1=\noexpand\iffalse}%
     \global\@if#1{false}\escapechar=\count@}


\newlinechar=`\^^J
\overfullrule=0pt




\let\itfam=\undefined

\let\bffam=\undefined

\count18=3


\chardef\sharps="19


\mathchardef\alpha="710B \mathchardef\beta="710C
\mathchardef\gamma="710D \mathchardef\delta="710E
\mathchardef\epsilon="710F \mathchardef\zeta="7110
\mathchardef\eta="7111 \mathchardef\theta="7112
\mathchardef\iota="7113 \mathchardef\kappa="7114
\mathchardef\lambda="7115 \mathchardef\mu="7116 \mathchardef\nu="7117
\mathchardef\xi="7118 \mathchardef\pi="7119 \mathchardef\rho="711A
\mathchardef\sigma="711B \mathchardef\tau="711C
\mathchardef\upsilon="711D \mathchardef\phi="711E
\mathchardef\chi="711F \mathchardef\psi="7120
\mathchardef\omega="7121 \mathchardef\varepsilon="7122
\mathchardef\vartheta="7123 \mathchardef\varpi="7124
\mathchardef\varrho="7125 \mathchardef\varsigma="7126
\mathchardef\varphi="7127 \mathchardef\imath="717B
\mathchardef\jmath="717C \mathchardef\ell="7160 \mathchardef\wp="717D
\mathchardef\partial="7140 \mathchardef\flat="715B
\mathchardef\natural="715C \mathchardef\sharp="715D



\def\angle{{\vbox{\ialign{$\m@th\scriptstyle##$\crcr
     \not\mathrel{\mkern14mu}\crcr
     \noalign{\nointerlineskip}
     \mkern2.5mu\leaders\hrule height.34\rp@\hfill\mkern2.5mu\crcr}}}}
\def\vdots{\vbox{\baselineskip4\rp@ \lineskiplimit\z@
     \kern6\rp@\hbox{.}\hbox{.}\hbox{.}}}
\def\ddots{\mathinner{\mkern1mu\raise7\rp@\vbox{\kern7\rp@\hbox{.}}\mkern2mu
     \raise4\rp@\hbox{.}\mkern2mu\raise\rp@\hbox{.}\mkern1mu}}
\def\overrightarrow#1{\vbox{\ialign{##\crcr
     \rightarrowfill\crcr
     \noalign{\kern-\rp@\nointerlineskip}
     $\hfil\displaystyle{#1}\hfil$\crcr}}}
\def\overleftarrow#1{\vbox{\ialign{##\crcr
     \leftarrowfill\crcr
     \noalign{\kern-\rp@\nointerlineskip}
     $\hfil\displaystyle{#1}\hfil$\crcr}}}
\def\overbrace#1{\mathop{\vbox{\ialign{##\crcr
     \noalign{\kern3\rp@}
     \downbracefill\crcr
     \noalign{\kern3\rp@\nointerlineskip}
     $\hfil\displaystyle{#1}\hfil$\crcr}}}\limits}
\def\underbrace#1{\mathop{\vtop{\ialign{##\crcr
     $\hfil\displaystyle{#1}\hfil$\crcr
     \noalign{\kern3\rp@\nointerlineskip}
     \upbracefill\crcr
     \noalign{\kern3\rp@}}}}\limits}
\def\big#1{{\hbox{$\left#1\vbox to8.5\rp@ {}\right.\n@space$}}}
\def\Big#1{{\hbox{$\left#1\vbox to11.5\rp@ {}\right.\n@space$}}}
\def\bigg#1{{\hbox{$\left#1\vbox to14.5\rp@ {}\right.\n@space$}}}
\def\Bigg#1{{\hbox{$\left#1\vbox to17.5\rp@ {}\right.\n@space$}}}
\def\@vereq#1#2{\lower.5\rp@\vbox{\baselineskip\z@skip\lineskip-.5\rp@
     \ialign{$\m@th#1\hfil##\hfil$\crcr#2\crcr=\crcr}}}
\def\rlh@#1{\vcenter{\hbox{\ooalign{\raise2\rp@
     \hbox{$#1\rightharpoonup$}\crcr
     $#1\leftharpoondown$}}}}
\def\bordermatrix#1{\begingroup\m@th
     \setbox\z@\vbox{%
          \def\cr{\crcr\noalign{\kern2\rp@\global\let\cr\endline}}%
          \ialign{$##$\hfil\kern2\rp@\kern\p@renwd
               &\thinspace\hfil$##$\hfil&&\quad\hfil$##$\hfil\crcr
               \omit\strut\hfil\crcr
               \noalign{\kern-\baselineskip}%
               #1\crcr\omit\strut\cr}}%
     \setbox\tw@\vbox{\unvcopy\z@\global\setbox\@ne\lastbox}%
     \setbox\tw@\hbox{\unhbox\@ne\unskip\global\setbox\@ne\lastbox}%
     \setbox\tw@\hbox{$\kern\wd\@ne\kern-\p@renwd\left(\kern-\wd\@ne
          \global\setbox\@ne\vbox{\box\@ne\kern2\rp@}%
          \vcenter{\kern-\ht\@ne\unvbox\z@\kern-\baselineskip}%
          \,\right)$}%
     \null\;\vbox{\kern\ht\@ne\box\tw@}\endgroup}
\def\endinsert{\egroup
     \if@mid\dimen@\ht\z@
          \advance\dimen@\dp\z@
          \advance\dimen@12\rp@
          \advance\dimen@\pagetotal
          \ifdim\dimen@>\pagegoal\@midfalse\p@gefalse\fi
     \fi
     \if@mid\bigskip\box\z@
          \bigbreak
     \else\insert\topins{\penalty100 \splittopskip\z@skip
               \splitmaxdepth\maxdimen\floatingpenalty\z@
               \ifp@ge\dimen@\dp\z@
                    \vbox to\vsize{\unvbox\z@\kern-\dimen@}%
               \else\box\z@\nobreak\bigskip
               \fi}%
     \fi
     \endgroup}


\def\cases#1{\left\{\,\vcenter{\m@th
     \ialign{$##\hfil$&\quad##\hfil\crcr#1\crcr}}\right.}
\def\matrix#1{\null\,\vcenter{\m@th
     \ialign{\hfil$##$\hfil&&\quad\hfil$##$\hfil\crcr
          \mathstrut\crcr
          \noalign{\kern-\baselineskip}
          #1\crcr
          \mathstrut\crcr
          \noalign{\kern-\baselineskip}}}\,}


\newif\ifraggedbottom

\def\raggedbottom{\ifraggedbottom\else
     \advance\topskip by\z@ plus60pt \raggedbottomtrue\fi}%
\def\normalbottom{\ifraggedbottom
     \advance\topskip by\z@ plus-60pt \raggedbottomfalse\fi}

\message{hacks,}


\toksdef\toks@i=1 \toksdef\toks@ii=2


\def\TeX{T\kern-.1667em \lower.5ex \hbox{E}\kern-.125em X\null}
\def\jyTeX{{\leavevmode
     \raise.587ex \hbox{\it\j}\kern-.1em \lower.048ex \hbox{\it y}\kern-.12em
     \TeX}}

\let\then=\iftrue
\def\ifnoarg#1\then{\def\hack@{#1}\ifx\hack@\empty}
\def\ifundefined#1\then{%
     \expandafter\ifx\csname\expandafter\blank\string#1\endcsname\relax}
\def\useif#1\then{\csname#1\endcsname}
\def\usename#1{\csname#1\endcsname}
\def\useafter#1#2{\expandafter#1\csname#2\endcsname}

\long\def\loop#1\repeat{\def\@iterate{#1\expandafter\@iterate\fi}\@iterate
     \let\@iterate=\relax}

\let\TeXend=\end
\def\begin#1{\begingroup\def\@@blockname{#1}\usename{begin#1}}
\def\end#1{\usename{end#1}\def\hack@{#1}%
     \ifx\@@blockname\hack@
          \endgroup
     \else\err@badgroup\hack@\@@blockname
     \fi}
\def\@@blockname{}

\def\defaultoption[#1]#2{%
     \def\hack@{\ifx\hack@ii[\toks@={#2}\else\toks@={#2[#1]}\fi\the\toks@}%
     \futurelet\hack@ii\hack@}

\def\markup#1{\let\@@marksf=\empty
     \ifhmode\edef\@@marksf{\spacefactor=\the\spacefactor\relax}\/\fi
     ${}^{\hbox{\subscriptfonts#1}}$\@@marksf}


\newtoks\shortyear
\newtoks\militaryhour
\newtoks\standardhour
\newtoks\minute
\newtoks\amorpm

\def\settime{\count@=\time\divide\count@ by60
     \militaryhour=\expandafter{\number\count@}%
     {\multiply\count@ by-60 \advance\count@ by\time
          \xdef\hack@{\ifnum\count@<10 0\fi\number\count@}}%
     \minute=\expandafter{\hack@}%
     \ifnum\count@<12
          \amorpm={am}
     \else\amorpm={pm}
          \ifnum\count@>12 \advance\count@ by-12 \fi
     \fi
     \standardhour=\expandafter{\number\count@}%
     \def\hack@19##1##2{\shortyear={##1##2}}%
          \expandafter\hack@\the\year}

\def\monthword#1{%
     \ifcase#1
          $\bullet$\err@badcountervalue{monthword}%
          \or January\or February\or March\or April\or May\or June%
          \or July\or August\or September\or October\or November\or December%
     \else$\bullet$\err@badcountervalue{monthword}%
     \fi}

\def\monthabbr#1{%
     \ifcase#1
          $\bullet$\err@badcountervalue{monthabbr}%
          \or Jan\or Feb\or Mar\or Apr\or May\or Jun%
          \or Jul\or Aug\or Sep\or Oct\or Nov\or Dec%
     \else$\bullet$\err@badcountervalue{monthabbr}%
     \fi}

\def\militarytime{\the\militaryhour:\the\minute}
\def\standardtime{\the\standardhour:\the\minute}


\def\@setnumstyle#1#2{\expandafter\global\expandafter\expandafter
     \expandafter\let\expandafter\expandafter
     \csname @\expandafter\blank\string#1style\endcsname
     \csname#2\endcsname}
\def\numstyle#1{\usename{@\expandafter\blank\string#1style}#1}
\def\ifblank#1\then{\useafter\ifx{@\expandafter\blank\string#1}\blank}

\def\blank#1{}

\def\Roman#1{\expandafter\uppercase\expandafter{\romannumeral#1}}
\def\alphabetic#1{%
     \ifcase#1
          $\bullet$\err@badcountervalue{alphabetic}%
          \or a\or b\or c\or d\or e\or f\or g\or h\or i\or j\or k\or l\or m%
          \or n\or o\or p\or q\or r\or s\or t\or u\or v\or w\or x\or y\or z%
     \else$\bullet$\err@badcountervalue{alphabetic}%
     \fi}
\def\Alphabetic#1{\expandafter\uppercase\expandafter{\alphabetic{#1}}}
\def\symbols#1{%
     \ifcase#1
          $\bullet$\err@badcountervalue{symbols}%
          \or*\or\dag\or\ddag\or\S\or$\|$%
          \or**\or\dag\dag\or\ddag\ddag\or\S\S\or$\|\|$%
     \else$\bullet$\err@badcountervalue{symbols}%
     \fi}


\catcode`\^^?=13 \def^^?{\relax}

\def\trimleading#1\to#2{\edef#2{#1}%
     \expandafter\@trimleading\expandafter#2#2^^?^^?}
\def\@trimleading#1#2#3^^?{\ifx#2^^?\def#1{}\else\def#1{#2#3}\fi}

\def\trimtrailing#1\to#2{\edef#2{#1}%
     \expandafter\@trimtrailing\expandafter#2#2^^? ^^?\relax}
\def\@trimtrailing#1#2 ^^?#3{\ifx#3\relax\toks@={}%
     \else\def#1{#2}\toks@={\trimtrailing#1\to#1}\fi
     \the\toks@}

\def\trim#1\to#2{\trimleading#1\to#2\trimtrailing#2\to#2}

\catcode`\^^?=15


\long\def\additemL#1\to#2{\toks@={\^^\{#1}}\toks@ii=\expandafter{#2}%
     \xdef#2{\the\toks@\the\toks@ii}}

\long\def\additemR#1\to#2{\toks@={\^^\{#1}}\toks@ii=\expandafter{#2}%
     \xdef#2{\the\toks@ii\the\toks@}}

\def\getitemL#1\to#2{\expandafter\@getitemL#1\hack@#1#2}
\def\@getitemL\^^\#1#2\hack@#3#4{\def#4{#1}\def#3{#2}}

\message{font macros,}


\newdimen\rp@
\newcount\@@sizeindex \@@sizeindex=0
\newcount\@@factori
\newcount\@@factorii
\newcount\@@factoriii
\newcount\@@factoriv

\countdef\maxfam=18
\newfam\itfam
\newfam\bffam
\newfam\bfsfam
\newfam\bmitfam

\def\@mathfontinit{\count@=4
     \loop\textfont\count@=\nullfont
          \scriptfont\count@=\nullfont
          \scriptscriptfont\count@=\nullfont
          \ifnum\count@<\maxfam\advance\count@ by\@ne
     \repeat}

\def\@fontstyleinit{%
     \def\it{\err@fontnotavailable\it}%
     \def\bf{\err@fontnotavailable\bf}%
     \def\bfs{\err@bfstobf}%
     \def\bmit{\err@fontnotavailable\bmit}%
     \def\sc{\err@fontnotavailable\sc}%
     \def\sl{\err@sltoit}%
     \def\ss{\err@fontnotavailable\ss}%
     \def\tt{\err@fontnotavailable\tt}}

\def\@parameterinit#1{\rm\rp@=.1em \@getscaling{#1}%
     \let\^^\=\@doscaling\scalingskipslist
     \setbox\strutbox=\hbox{\vrule
          height.708\baselineskip depth.292\baselineskip width\z@}}

\def\@getfactor#1#2#3#4{\@@factori=#1 \@@factorii=#2
     \@@factoriii=#3 \@@factoriv=#4}

\def\@getscaling#1{\count@=#1 \advance\count@ by-\@@sizeindex\@@sizeindex=#1
     \ifnum\count@<0
          \let\@mulordiv=\divide
          \let\@divormul=\multiply
          \multiply\count@ by\m@ne
     \else\let\@mulordiv=\multiply
          \let\@divormul=\divide
     \fi
     \edef\@@scratcha{\ifcase\count@                {1}{1}{1}{1}\or
          {1}{7}{23}{3}\or     {2}{5}{3}{1}\or      {9}{89}{13}{1}\or
          {6}{25}{6}{1}\or     {8}{71}{14}{1}\or    {6}{25}{36}{5}\or
          {1}{7}{53}{4}\or     {12}{125}{108}{5}\or {3}{14}{53}{5}\or
          {6}{41}{17}{1}\or    {13}{31}{13}{2}\or   {9}{107}{71}{2}\or
          {11}{139}{124}{3}\or {1}{6}{43}{2}\or     {10}{107}{42}{1}\or
          {1}{5}{43}{2}\or     {5}{69}{65}{1}\or    {11}{97}{91}{2}\fi}%
     \expandafter\@getfactor\@@scratcha}

\def\@doscaling#1{\@mulordiv#1by\@@factori\@divormul#1by\@@factorii
     \@mulordiv#1by\@@factoriii\@divormul#1by\@@factoriv}


\newskip\headskip
\newskip\footskip

\def\typesize=#1pt{\count@=#1 \advance\count@ by-10
     \ifcase\count@
          \@setsizex\or\err@badtypesize\or
          \@setsizexii\or\err@badtypesize\or
          \@setsizexiv
     \else\err@badtypesize
     \fi}

\def\@setsizex{\getixpt
     \def\subsubscriptfonts{\vpt}%
          \def\subsubscriptsize{\vpt\@parameterinit{-8}}%
     \def\subscriptfonts{\viipt}\def\subscriptsize{\viipt\@parameterinit{-4}}%
     \def\footnotefonts{\viiipt}\def\footnotesize{\viiipt\@parameterinit{-2}}%
     \def\smallfonts{\ixpt}\def\smallsize{\ixpt\@parameterinit{-1}}%
     \def\normalfonts{\xpt}\def\normalsize{\xpt\@parameterinit{0}}%
     \def\bigfonts{\xiipt}\def\bigsize{\xiipt\@parameterinit{2}}%
     \def\Bigfonts{\xivpt}\def\Bigsize{\xivpt\@parameterinit{4}}%
     \def\biggfonts{\xviipt}\def\biggsize{\xviipt\@parameterinit{6}}%
     \def\Biggfonts{\xxipt}\def\Biggsize{\xxipt\@parameterinit{8}}%
     \def\tinyfonts{\vpt}\def\tinysize{\vpt\@parameterinit{-8}}%
     \def\HUGEFONTS{\xxvpt}\def\HUGESIZE{\xxvpt\@parameterinit{10}}%
     \normalsize\fixedskipslist}

\def\@setsizexii{\getxipt
     \def\subsubscriptfonts{\vipt}%
          \def\subsubscriptsize{\vipt\@parameterinit{-6}}%
     \def\subscriptfonts{\viiipt}%
          \def\subscriptsize{\viiipt\@parameterinit{-2}}%
     \def\footnotefonts{\xpt}\def\footnotesize{\xpt\@parameterinit{0}}%
     \def\smallfonts{\xipt}\def\smallsize{\xipt\@parameterinit{1}}%
     \def\normalfonts{\xiipt}\def\normalsize{\xiipt\@parameterinit{2}}%
     \def\bigfonts{\xivpt}\def\bigsize{\xivpt\@parameterinit{4}}%
     \def\Bigfonts{\xviipt}\def\Bigsize{\xviipt\@parameterinit{6}}%
     \def\biggfonts{\xxipt}\def\biggsize{\xxipt\@parameterinit{8}}%
     \def\Biggfonts{\xxvpt}\def\Biggsize{\xxvpt\@parameterinit{10}}%
     \def\tinyfonts{\vpt}\def\tinysize{\vpt\@parameterinit{-8}}%
     \def\HUGEFONTS{\xxvpt}\def\HUGESIZE{\xxvpt\@parameterinit{10}}%
     \normalsize\fixedskipslist}

\def\@setsizexiv{\getxiiipt
     \def\subsubscriptfonts{\viipt}%
          \def\subsubscriptsize{\viipt\@parameterinit{-4}}%
     \def\subscriptfonts{\xpt}\def\subscriptsize{\xpt\@parameterinit{0}}%
     \def\footnotefonts{\xiipt}\def\footnotesize{\xiipt\@parameterinit{2}}%
     \def\smallfonts{\xiiipt}\def\smallsize{\xiiipt\@parameterinit{3}}%
     \def\normalfonts{\xivpt}\def\normalsize{\xivpt\@parameterinit{4}}%
     \def\bigfonts{\xviipt}\def\bigsize{\xviipt\@parameterinit{6}}%
     \def\Bigfonts{\xxipt}\def\Bigsize{\xxipt\@parameterinit{8}}%
     \def\biggfonts{\xxvpt}\def\biggsize{\xxvpt\@parameterinit{10}}%
     \def\Biggfonts{\err@sizetoolarge\Biggfonts\HUGEFONTS}%
          \def\Biggsize{\err@sizetoolarge\Biggsize\HUGESIZE}%
     \def\tinyfonts{\vpt}\def\tinysize{\vpt\@parameterinit{-8}}%
     \def\HUGEFONTS{\xxvpt}\def\HUGESIZE{\xxvpt\@parameterinit{10}}%
     \normalsize\fixedskipslist}

\def\subsubscriptfonts{\vpt} \def\subsubscriptsize{\vpt\@parameterinit{-8}}
\def\subscriptfonts{\viipt}  \def\subscriptsize{\viipt\@parameterinit{-4}}
\def\footnotefonts{\viiipt}  \def\footnotesize{\viiipt\@parameterinit{-2}}
\def\smallfonts{\err@sizenotavailable\smallfonts}
                             \def\smallsize{\ixpt\@parameterinit{-1}}
\def\normalfonts{\xpt}       \def\normalsize{\xpt\@parameterinit{0}}
\def\bigfonts{\xiipt}        \def\bigsize{\xiipt\@parameterinit{2}}
\def\Bigfonts{\xivpt}        \def\Bigsize{\xivpt\@parameterinit{4}}
\def\biggfonts{\xviipt}      \def\biggsize{\xviipt\@parameterinit{6}}
\def\Biggfonts{\xxipt}       \def\Biggsize{\xxipt\@parameterinit{8}}
\def\tinyfonts{\vpt}         \def\tinysize{\vpt\@parameterinit{-8}}
\def\HUGEFONTS{\xxvpt}       \def\HUGESIZE{\xxvpt\@parameterinit{10}}

\message{document layout,}


\newtoks\everyoutput \everyoutput={}
\newdimen\depthofpage
\newcount\pagenum \pagenum=0

\newdimen\oddtopmargin  \newdimen\eventopmargin
\newdimen\oddleftmargin \newdimen\evenleftmargin
\newtoks\oddhead        \newtoks\evenhead
\newtoks\oddfoot        \newtoks\evenfoot

\def\topmargin{\afterassignment\@seteventop\oddtopmargin}
\def\leftmargin{\afterassignment\@setevenleft\oddleftmargin}
\def\head{\afterassignment\@setevenhead\oddhead}
\def\foot{\afterassignment\@setevenfoot\oddfoot}

\def\@seteventop{\eventopmargin=\oddtopmargin}
\def\@setevenleft{\evenleftmargin=\oddleftmargin}
\def\@setevenhead{\evenhead=\oddhead}
\def\@setevenfoot{\evenfoot=\oddfoot}

\def\pagenumstyle#1{\@setnumstyle\pagenum{#1}}

\newif\ifdraft
\def\draft{\drafttrue\leftmargin=.5in \overfullrule=5pt }

\def\outputstyle#1{\global\expandafter\let\expandafter
          \@outputstyle\csname#1output\endcsname
     \usename{#1setup}}

\output={\@outputstyle}

\def\normaloutput{\the\everyoutput
     \global\advance\pagenum by\@ne
     \ifodd\pagenum
          \voffset=\oddtopmargin \hoffset=\oddleftmargin
     \else\voffset=\eventopmargin \hoffset=\evenleftmargin
     \fi
     \advance\voffset by-1in  \advance\hoffset by-1in
     \count0=\pagenum
     \expandafter\shipout\pagebox
     \ifnum\outputpenalty>-\@MM\else\dosupereject\fi}

\newdimen\fullhsize
\newbox\leftpage
\newcount\leftpagenum
\newcount\outputpagenum \outputpagenum=0
\let\leftorright=L

\def\twoupoutput{\the\everyoutput
     \global\advance\pagenum by\@ne
     \if L\leftorright
          \global\setbox\leftpage=\leftline{\pagebox}%
          \global\leftpagenum=\pagenum
          \global\let\leftorright=R%
     \else\global\advance\outputpagenum by\@ne
          \ifodd\outputpagenum
               \voffset=\oddtopmargin \hoffset=\oddleftmargin
          \else\voffset=\eventopmargin \hoffset=\evenleftmargin
          \fi
          \advance\voffset by-1in  \advance\hoffset by-1in
          \count0=\leftpagenum \count1=\pagenum
          \shipout\vbox{\hbox to\fullhsize
               {\box\leftpage\hfil\leftline{\pagebox}}}%
          \global\let\leftorright=L%
     \fi
     \ifnum\outputpenalty>-\@MM
     \else\dosupereject
          \if R\leftorright
               \globaldefs=\@ne\head={\hfil}\foot={\hfil}\globaldefs=\z@
               \null\newpage
          \fi
     \fi}

\def\pagebox{\vbox{\makeheadline\pagebody\makefootline}}

\def\makeheadline{%
     \vbox to\z@{\baselinestretch=\@m
          \vskip\topskip\vskip-.708\baselineskip\vskip-\headskip
          \line{\vbox to\ht\strutbox{}%
               \ifodd\pagenum\the\oddhead\else\the\evenhead\fi}%
          \vss}%
     \nointerlineskip}

\def\pagebody{\vbox to\vsize{%
     \boxmaxdepth\maxdepth
     \ifvoid\topins\else\unvbox\topins\fi
     \depthofpage=\dp255
     \unvbox255
     \ifraggedbottom\kern-\depthofpage\vfil\fi
     \ifvoid\footins
     \else\vskip\skip\footins
          \footnoterule
          \unvbox\footins
          \vskip-\footnoteskip
     \fi}}

\def\makefootline{\baselineskip=\footskip
     \line{\ifodd\pagenum\the\oddfoot\else\the\evenfoot\fi}}


\newskip\abovechapterskip
\newskip\belowchapterskip
\newskip\abovesectionskip
\newskip\belowsectionskip
\newskip\abovesubsectionskip
\newskip\belowsubsectionskip

\def\chapterstyle#1{\global\expandafter\let\expandafter\@chapterstyle
     \csname#1text\endcsname}
\def\sectionstyle#1{\global\expandafter\let\expandafter\@sectionstyle
     \csname#1text\endcsname}
\def\subsectionstyle#1{\global\expandafter\let\expandafter\@subsectionstyle
     \csname#1text\endcsname}

\def\chapter#1{%
     \ifdim\lastskip=17sp \else\chapterbreak\vskip\abovechapterskip\fi
     \@chapterstyle{\ifblank\chapternumstyle\then
          \else\newchapternum=\next\chapternumformat\ \fi#1}%
     \nobreak\vskip\belowchapterskip\vskip17sp }

\def\section#1{%
     \ifdim\lastskip=17sp \else\sectionbreak\vskip\abovesectionskip\fi
     \@sectionstyle{\ifblank\sectionnumstyle\then
          \else\newsectionnum=\next\sectionnumformat\ \fi#1}%
     \nobreak\vskip\belowsectionskip\vskip17sp }

\def\subsection#1{%
     \ifdim\lastskip=17sp \else\subsectionbreak\vskip\abovesubsectionskip\fi
     \@subsectionstyle{\ifblank\subsectionnumstyle\then
          \else\newsubsectionnum=\next\subsectionnumformat\ \fi#1}%
     \nobreak\vskip\belowsubsectionskip\vskip17sp }


\let\TeXunderline=\underline
\let\TeXoverline=\overline
\def\underline#1{\relax\ifmmode\TeXunderline{#1}\else
     $\TeXunderline{\hbox{#1}}$\fi}
\def\overline#1{\relax\ifmmode\TeXoverline{#1}\else
     $\TeXoverline{\hbox{#1}}$\fi}

\def\baselinestretch{\afterassignment\@baselinestretch\count@}
\def\@baselinestretch{\baselineskip=\normalbaselineskip
     \divide\baselineskip by\@m\baselineskip=\count@\baselineskip
     \setbox\strutbox=\hbox{\vrule
          height.708\baselineskip depth.292\baselineskip width\z@}%
     \bigskipamount=\the\baselineskip
          plus.25\baselineskip minus.25\baselineskip
     \medskipamount=.5\baselineskip
          plus.125\baselineskip minus.125\baselineskip
     \smallskipamount=.25\baselineskip
          plus.0625\baselineskip minus.0625\baselineskip}

\def\\{\ifhmode\ifnum\lastpenalty=-\@M\else\hfil\penalty-\@M\fi\fi
     \ignorespaces}
\def\newpage{\vfil\break}

\def\lefttext#1{\par{\@text\leftskip=\z@\rightskip=\centering
     \noindent#1\par}}
\def\righttext#1{\par{\@text\leftskip=\centering\rightskip=\z@
     \noindent#1\par}}
\def\centertext#1{\par{\@text\leftskip=\centering\rightskip=\centering
     \noindent#1\par}}
\def\@text{\parindent=\z@ \parfillskip=\z@ \everypar={}%
     \spaceskip=.3333em \xspaceskip=.5em
     \def\\{\ifhmode\ifnum\lastpenalty=-\@M\else\penalty-\@M\fi\fi
          \ignorespaces}}

\def\beginleft{\par\@text\leftskip=\z@ \rightskip=\centering}
     
\def\beginright{\par\@text\leftskip=\centering\rightskip=\z@ }
     
\def\begincenter{\par\@text\leftskip=\centering\rightskip=\centering}

\def\beginnarrow{\defaultoption[\parindent]\@beginnarrow}
\def\@beginnarrow[#1]{\par\advance\leftskip by#1\advance\rightskip by#1}

\begingroup
\catcode`\[=1 \catcode`\{=11 \gdef\beginignore[\endgroup\bgroup
     \catcode`\e=0 \catcode`\\=12 \catcode`\{=11 \catcode`\f=12 \let\or=\relax
     \let\nd{ignor=\fi \let\}=\egroup
     \iffalse}
\endgroup

\long\def\marginnote#1{\leavevmode
     \edef\@marginsf{\spacefactor=\the\spacefactor\relax}%
     \ifdraft\strut\vadjust{%
          \hbox to\z@{\hskip\hsize\hskip.1in
               \vbox to\z@{\vskip-\dp\strutbox
                    \marginnoteformat
                    \vskip-\ht\strutbox
                    \noindent\strut#1\par
                    \vss}%
               \hss}}%
     \fi
     \@marginsf}


\newtoks\everybye \everybye={\par\vfil}
\outer\def\bye{\the\everybye
     \footnotecheck
     \prelabelcheck
     \streamcheck
     \supereject
     \TeXend}

\message{footnotes,}

\newcount\footnotenum \footnotenum=0
\newskip\footnoteskip
\let\@footnotelist=\empty

\def\footnotenumstyle#1{\@setnumstyle\footnotenum{#1}%
     \useafter\ifx{@footnotenumstyle}\symbols
          \global\let\@footup=\empty
     \else\global\let\@footup=\markup
     \fi}

\def\footnote{\footnotecheck\defaultoption[]\@footnote}
\def\@footnote[#1]{\@footnotemark[#1]\@footnotetext}

\def\footnotemark{\defaultoption[]\@footnotemark}
\def\@footnotemark[#1]{\let\@footsf=\empty
     \ifhmode\edef\@footsf{\spacefactor=\the\spacefactor\relax}\/\fi
     \ifnoarg#1\then
          \global\advance\footnotenum by\@ne
          \@footup{\footnotenumformat}%
          \edef\@@foota{\footnotenum=\the\footnotenum\relax}%
          \expandafter\additemR\expandafter\@footup\expandafter
               {\@@foota\footnotenumformat}\to\@footnotelist
          \global\let\@footnotelist=\@footnotelist
     \else\markup{#1}%
          \additemR\markup{#1}\to\@footnotelist
          \global\let\@footnotelist=\@footnotelist
     \fi
     \@footsf}

\def\footnotetext{%
     \ifx\@footnotelist\empty\err@extrafootnotetext\else\@footnotetext\fi}
\def\@footnotetext{%
     \getitemL\@footnotelist\to\@@foota
     \global\let\@footnotelist=\@footnotelist
     \insert\footins\bgroup
     \footnoteformat
     \splittopskip=\ht\strutbox\splitmaxdepth=\dp\strutbox
     \interlinepenalty=\interfootnotelinepenalty\floatingpenalty=\@MM
     \noindent\llap{\@@foota}\strut
     \bgroup\aftergroup\@footnoteend
     \let\@@scratcha=}
\def\@footnoteend{\strut\par\vskip\footnoteskip\egroup}

\def\footnoterule{\normalfonts
     \kern-.3em \hrule width2in height.04em \kern .26em }

\def\footnotecheck{%
     \ifx\@footnotelist\empty
     \else\err@extrafootnotemark
          \global\let\@footnotelist=\empty
     \fi}

\message{labels,}

\let\@@labeldef=\xdef
\newif\if@labelfile
\newwrite\@labelfile
\let\@prelabellist=\empty

\def\label#1#2{\trim#1\to\@@labarg\edef\@@labtext{#2}%
     \edef\@@labname{lab@\@@labarg}%
     \useafter\ifundefined\@@labname\then\else\@yeslab\fi
     \useafter\@@labeldef\@@labname{#2}%
     \ifstreaming
          \expandafter\toks@\expandafter\expandafter\expandafter
               {\csname\@@labname\endcsname}%
          \immediate\write\streamout{\noexpand\label{\@@labarg}{\the\toks@}}%
     \fi}
\def\@yeslab{%
     \useafter\ifundefined{if\@@labname}\then
          \err@labelredef\@@labarg
     \else\useif{if\@@labname}\then
               \err@labelredef\@@labarg
          \else\global\usename{\@@labname true}%
               \useafter\ifundefined{pre\@@labname}\then
               \else\useafter\ifx{pre\@@labname}\@@labtext
                    \else\err@badlabelmatch\@@labarg
                    \fi
               \fi
               \if@labelfile
               \else\global\@labelfiletrue
                    \immediate\write\sixt@@n{--> Creating file \jobname.lab}%
                    \immediate\openout\@labelfile=\jobname.lab
               \fi
               \immediate\write\@labelfile
                    {\noexpand\prelabel{\@@labarg}{\@@labtext}}%
          \fi
     \fi}

\def\putlab#1{\trim#1\to\@@labarg\edef\@@labname{lab@\@@labarg}%
     \useafter\ifundefined\@@labname\then\@nolab\else\usename\@@labname\fi}
\def\@nolab{%
     \useafter\ifundefined{pre\@@labname}\then
          \undefinedlabelformat
          \err@needlabel\@@labarg
          \useafter\xdef\@@labname{\undefinedlabelformat}%
     \else\usename{pre\@@labname}%
          \useafter\xdef\@@labname{\usename{pre\@@labname}}%
     \fi
     \useafter\newif{if\@@labname}%
     \expandafter\additemR\@@labarg\to\@prelabellist}

\def\prelabel#1{\useafter\gdef{prelab@#1}}

\def\ifundefinedlabel#1\then{%
     \expandafter\ifx\csname lab@#1\endcsname\relax}
\def\useiflab#1\then{\csname iflab@#1\endcsname}

\def\prelabelcheck{{%
     \def\^^\##1{\useiflab{##1}\then\else\err@undefinedlabel{##1}\fi}%
     \@prelabellist}}

\message{equation numbering,}

\newcount\chapternum
\newcount\sectionnum
\newcount\subsectionnum
\newcount\equationnum
\newcount\subequationnum
\newcount\figurenum
\newcount\subfigurenum
\newcount\tablenum
\newcount\subtablenum

\newif\if@subeqncount
\newif\if@subfigcount
\newif\if@subtblcount

\def\newchapternum{\newsectionnum=\z@\@resetnum\chapternum}
\def\newsectionnum{\newsubsectionnum=\z@\@resetnum\sectionnum}
\def\newsubsectionnum{\newequationnum=\z@\newfigurenum=\z@\newtablenum=\z@
     \@resetnum\subsectionnum}
\def\newequationnum{\newsubequationnum=\z@\@resetnum\equationnum}
\def\newsubequationnum{\@resetnum\subequationnum}
\def\newfigurenum{\newsubfigurenum=\z@\@resetnum\figurenum}
\def\newsubfigurenum{\@resetnum\subfigurenum}
\def\newtablenum{\newsubtablenum=\z@\@resetnum\tablenum}
\def\newsubtablenum{\@resetnum\subtablenum}

\def\@resetnum#1{\global\advance#1by1 \edef\next{\the#1\relax}\global#1}

\newchapternum=0

\def\chapternumstyle#1{\@setnumstyle\chapternum{#1}}
\def\sectionnumstyle#1{\@setnumstyle\sectionnum{#1}}
\def\subsectionnumstyle#1{\@setnumstyle\subsectionnum{#1}}
\def\equationnumstyle#1{\@setnumstyle\equationnum{#1}}
\def\subequationnumstyle#1{\@setnumstyle\subequationnum{#1}%
     \ifblank\subequationnumstyle\then\global\@subeqncountfalse\fi
     \ignorespaces}
\def\figurenumstyle#1{\@setnumstyle\figurenum{#1}}
\def\subfigurenumstyle#1{\@setnumstyle\subfigurenum{#1}%
     \ifblank\subfigurenumstyle\then\global\@subfigcountfalse\fi
     \ignorespaces}
\def\tablenumstyle#1{\@setnumstyle\tablenum{#1}}
\def\subtablenumstyle#1{\@setnumstyle\subtablenum{#1}%
     \ifblank\subtablenumstyle\then\global\@subtblcountfalse\fi
     \ignorespaces}

\def\eqnlabel#1{%
     \if@subeqncount
          \newsubequationnum=\next
     \else\newequationnum=\next
          \ifblank\subequationnumstyle\then
          \else\global\@subeqncounttrue
               \newsubequationnum=\@ne
          \fi
     \fi
     \label{#1}{\puteqnformat}(\puteqn{#1})%
     \ifdraft\rlap{\hskip.1in{\tt#1}}\fi}

\let\puteqn=\putlab

\def\equation#1#2{\useafter\gdef{eqn@#1}{#2\eqno\eqnlabel{#1}}}
\def\Equation#1{\useafter\gdef{eqn@#1}}

\def\putequation#1{\useafter\ifundefined{eqn@#1}\then
     \err@undefinedeqn{#1}\else\usename{eqn@#1}\fi}

\def\eqnseriesstyle#1{\gdef\@eqnseriesstyle{#1}}
\def\begineqnseries{\subequationnumstyle{\@eqnseriesstyle}%
     \defaultoption[]\@begineqnseries}
\def\@begineqnseries[#1]{\edef\@@eqnname{#1}}
\def\endeqnseries{\subequationnumstyle{blank}%
     \expandafter\ifnoarg\@@eqnname\then
     \else\label\@@eqnname{\puteqnformat}%
     \fi
     \aftergroup\ignorespaces}

\def\figlabel#1{%
     \if@subfigcount
          \newsubfigurenum=\next
     \else\newfigurenum=\next
          \ifblank\subfigurenumstyle\then
          \else\global\@subfigcounttrue
               \newsubfigurenum=\@ne
          \fi
     \fi
     \label{#1}{\putfigformat}\putfig{#1}%
     {\def\marginnoteformat{\tt}\marginnote{#1}}}

\let\putfig=\putlab

\def\figseriesstyle#1{\gdef\@figseriesstyle{#1}}
\def\beginfigseries{\subfigurenumstyle{\@figseriesstyle}%
     \defaultoption[]\@beginfigseries}
\def\@beginfigseries[#1]{\edef\@@figname{#1}}
\def\endfigseries{\subfigurenumstyle{blank}%
     \expandafter\ifnoarg\@@figname\then
     \else\label\@@figname{\putfigformat}%
     \fi
     \aftergroup\ignorespaces}

\def\tbllabel#1{%
     \if@subtblcount
          \newsubtablenum=\next
     \else\newtablenum=\next
          \ifblank\subtablenumstyle\then
          \else\global\@subtblcounttrue
               \newsubtablenum=\@ne
          \fi
     \fi
     \label{#1}{\puttblformat}\puttbl{#1}%
     {\def\marginnoteformat{\tt}\marginnote{#1}}}

\let\puttbl=\putlab

\def\tblseriesstyle#1{\gdef\@tblseriesstyle{#1}}
\def\begintblseries{\subtablenumstyle{\@tblseriesstyle}%
     \defaultoption[]\@begintblseries}
\def\@begintblseries[#1]{\edef\@@tblname{#1}}
\def\endtblseries{\subtablenumstyle{blank}%
     \expandafter\ifnoarg\@@tblname\then
     \else\label\@@tblname{\puttblformat}%
     \fi
     \aftergroup\ignorespaces}

\message{reference numbering,}

\newcount\referencenum \referencenum=0
\newcount\@@prerefcount \@@prerefcount=0
\newcount\@@thisref
\newcount\@@lastref
\newcount\@@loopref
\newcount\@@refseq
\newdimen\refnumindent
\let\@undefreflist=\empty

\def\referencenumstyle#1{\@setnumstyle\referencenum{#1}}

\def\referencestyle#1{\usename{@ref#1}}

\def\@refsequential{%
     \gdef\@refpredef##1{\global\advance\referencenum by\@ne
          \let\^^\=0\label{##1}{\^^\{\the\referencenum}}%
          \useafter\gdef{ref@\the\referencenum}{{##1}{\undefinedlabelformat}}}%
     \gdef\@reference##1##2{%
          \ifundefinedlabel##1\then
          \else\def\^^\####1{\global\@@thisref=####1\relax}\putlab{##1}%
               \useafter\gdef{ref@\the\@@thisref}{{##1}{##2}}%
          \fi}%
     \gdef\endputreferences{%
          \loop\ifnum\@@loopref<\referencenum
                    \advance\@@loopref by\@ne
                    \expandafter\expandafter\expandafter\@printreference
                         \csname ref@\the\@@loopref\endcsname
          \repeat
          \par}}

\def\@refpreordered{%
     \gdef\@refpredef##1{\global\advance\referencenum by\@ne
          \additemR##1\to\@undefreflist}%
     \gdef\@reference##1##2{%
          \ifundefinedlabel##1\then
          \else\global\advance\@@loopref by\@ne
               {\let\^^\=0\label{##1}{\^^\{\the\@@loopref}}}%
               \@printreference{##1}{##2}%
          \fi}
     \gdef\endputreferences{%
          \def\^^\####1{\useiflab{####1}\then
               \else\reference{####1}{\undefinedlabelformat}\fi}%
          \@undefreflist
          \par}}

\def\beginprereferences{\par
     \def\reference##1##2{\global\advance\referencenum by1\@ne
          \let\^^\=0\label{##1}{\^^\{\the\referencenum}}%
          \useafter\gdef{ref@\the\referencenum}{{##1}{##2}}}}
\def\endprereferences{\global\@@prerefcount=\the\referencenum\par}

\def\beginputreferences{\par
     \refnumindent=\z@\@@loopref=\z@
     \loop\ifnum\@@loopref<\referencenum
               \advance\@@loopref by\@ne
               \setbox\z@=\hbox{\referencenum=\@@loopref
                    \referencenumformat\enskip}%
               \ifdim\wd\z@>\refnumindent\refnumindent=\wd\z@\fi
     \repeat
     \putreferenceformat
     \@@loopref=\z@
     \loop\ifnum\@@loopref<\@@prerefcount
               \advance\@@loopref by\@ne
               \expandafter\expandafter\expandafter\@printreference
                    \csname ref@\the\@@loopref\endcsname
     \repeat
     \let\reference=\@reference}

\def\@printreference#1#2{\ifx#2\undefinedlabelformat\err@undefinedref{#1}\fi
     \noindent\ifdraft\rlap{\hskip\hsize\hskip.1in \tt#1}\fi
     \llap{\referencenum=\@@loopref\referencenumformat\enskip}#2\par}

\def\reference#1#2{{\par\refnumindent=\z@\putreferenceformat\noindent#2\par}}

\def\putref#1{\trim#1\to\@@refarg
     \expandafter\ifnoarg\@@refarg\then
          \toks@={\relax}%
     \else\@@lastref=-\@m\def\@@refsep{}\def\@more{\@nextref}%
          \toks@={\@nextref#1,,}%
     \fi\the\toks@}
\def\@nextref#1,{\trim#1\to\@@refarg
     \expandafter\ifnoarg\@@refarg\then
          \let\@more=\relax
     \else\ifundefinedlabel\@@refarg\then
               \expandafter\@refpredef\expandafter{\@@refarg}%
          \fi
          \def\^^\##1{\global\@@thisref=##1\relax}%
          \global\@@thisref=\m@ne
          \setbox\z@=\hbox{\putlab\@@refarg}%
     \fi
     \advance\@@lastref by\@ne
     \ifnum\@@lastref=\@@thisref\advance\@@refseq by\@ne\else\@@refseq=\@ne\fi
     \ifnum\@@lastref<\z@
     \else\ifnum\@@refseq<\thr@@
               \@@refsep\def\@@refsep{,}%
               \ifnum\@@lastref>\z@
                    \advance\@@lastref by\m@ne
                    {\referencenum=\@@lastref\putrefformat}%
               \else\undefinedlabelformat
               \fi
          \else\def\@@refsep{--}%
          \fi
     \fi
     \@@lastref=\@@thisref
     \@more}

\message{streaming,}

\newif\ifstreaming

\def\streamto{\defaultoption[\jobname]\@streamto}
\def\@streamto[#1]{\global\streamingtrue
     \immediate\write\sixt@@n{--> Streaming to #1.str}%
     \newwrite\streamout\immediate\openout\streamout=#1.str }

\def\streamfrom{\defaultoption[\jobname]\@streamfrom}
\def\@streamfrom[#1]{\newread\streamin\openin\streamin=#1.str
     \ifeof\streamin
          \expandafter\err@nostream\expandafter{#1.str}%
     \else\immediate\write\sixt@@n{--> Streaming from #1.str}%
          \let\@@labeldef=\gdef
          \ifstreaming
               \edef\@elc{\endlinechar=\the\endlinechar}%
               \endlinechar=\m@ne
               \loop\read\streamin to\@@scratcha
                    \ifeof\streamin
                         \streamingfalse
                    \else\toks@=\expandafter{\@@scratcha}%
                         \immediate\write\streamout{\the\toks@}%
                    \fi
                    \ifstreaming
               \repeat
               \@elc
               \input #1.str
               \streamingtrue
          \else\input #1.str
          \fi
          \let\@@labeldef=\xdef
     \fi}

\def\streamcheck{\ifstreaming
     \immediate\write\streamout{\pagenum=\the\pagenum}%
     \immediate\write\streamout{\footnotenum=\the\footnotenum}%
     \immediate\write\streamout{\referencenum=\the\referencenum}%
     \immediate\write\streamout{\chapternum=\the\chapternum}%
     \immediate\write\streamout{\sectionnum=\the\sectionnum}%
     \immediate\write\streamout{\subsectionnum=\the\subsectionnum}%
     \immediate\write\streamout{\equationnum=\the\equationnum}%
     \immediate\write\streamout{\subequationnum=\the\subequationnum}%
     \immediate\write\streamout{\figurenum=\the\figurenum}%
     \immediate\write\streamout{\subfigurenum=\the\subfigurenum}%
     \immediate\write\streamout{\tablenum=\the\tablenum}%
     \immediate\write\streamout{\subtablenum=\the\subtablenum}%
     \immediate\closeout\streamout
     \fi}


\def\err@badtypesize{%
     \errhelp={The limited availability of certain fonts requires^^J%
          that the base type size be 10pt, 12pt, or 14pt.^^J}%
     \errmessage{--> Illegal base type size}}

\def\err@badsizechange{\immediate\write\sixt@@n
     {--> Size change not allowed in math mode, ignored}}

\def\err@sizetoolarge#1{\immediate\write\sixt@@n
     {--> \noexpand#1 too big, substituting HUGE}}

\def\err@sizenotavailable#1{\immediate\write\sixt@@n
     {--> Size not available, \noexpand#1 ignored}}

\def\err@fontnotavailable#1{\immediate\write\sixt@@n
     {--> Font not available, \noexpand#1 ignored}}

\def\err@sltoit{\immediate\write\sixt@@n
     {--> Style \noexpand\sl not available, substituting \noexpand\it}%
     \it}

\def\err@bfstobf{\immediate\write\sixt@@n
     {--> Style \noexpand\bfs not available, substituting \noexpand\bf}%
     \bf}

\def\err@badgroup#1#2{%
     \errhelp={The block you have just tried to close was not the one^^J%
          most recently opened.^^J}%
     \errmessage{--> \noexpand\end{#1} doesn't match \noexpand\begin{#2}}}

\def\err@badcountervalue#1{\immediate\write\sixt@@n
     {--> Counter (#1) out of bounds}}

\def\err@extrafootnotemark{\immediate\write\sixt@@n
     {--> \noexpand\footnotemark command
          has no corresponding \noexpand\footnotetext}}

\def\err@extrafootnotetext{%
     \errhelp{You have given a \noexpand\footnotetext command without first
          specifying^^Ja \noexpand\footnotemark.^^J}%
     \errmessage{--> \noexpand\footnotetext command has no corresponding
          \noexpand\footnotemark}}

\def\err@labelredef#1{\immediate\write\sixt@@n
     {--> Label "#1" redefined}}

\def\err@badlabelmatch#1{\immediate\write\sixt@@n
     {--> Definition of label "#1" doesn't match value in \jobname.lab}}

\def\err@needlabel#1{\immediate\write\sixt@@n
     {--> Label "#1" cited before its definition}}

\def\err@undefinedlabel#1{\immediate\write\sixt@@n
     {--> Label "#1" cited but never defined}}

\def\err@undefinedeqn#1{\immediate\write\sixt@@n
     {--> Equation "#1" not defined}}

\def\err@undefinedref#1{\immediate\write\sixt@@n
     {--> Reference "#1" not defined}}

\def\err@nostream#1{%
     \errhelp={You have tried to input a stream file that doesn't exist.^^J}%
     \errmessage{--> Stream file #1 not found}}

\message{jyTeX initialization}

\everyjob{\immediate\write16{--> jyTeX version \fmtversion}%
     \edef\@@jobname{\jobname}%
     \edef\jobname{\@@jobname}%
     \settime
     \openin0=\jobname.lab
     \ifeof0
     \else\closein0
          \immediate\write16{--> Getting labels from file \jobname.lab}%
          \input\jobname.lab
     \fi}


\def\fixedskipslist{%
     \^^\{\topskip}%
     \^^\{\splittopskip}%
     \^^\{\maxdepth}%
     \^^\{\skip\topins}%
     \^^\{\skip\footins}%
     \^^\{\headskip}%
     \^^\{\footskip}}

\def\scalingskipslist{%
     \^^\{\p@renwd}%
     \^^\{\delimitershortfall}%
     \^^\{\nulldelimiterspace}%
     \^^\{\scriptspace}%
     \^^\{\jot}%
     \^^\{\normalbaselineskip}%
     \^^\{\normallineskip}%
     \^^\{\normallineskiplimit}%
     \^^\{\baselineskip}%
     \^^\{\lineskip}%
     \^^\{\lineskiplimit}%
     \^^\{\bigskipamount}%
     \^^\{\medskipamount}%
     \^^\{\smallskipamount}%
     \^^\{\parskip}%
     \^^\{\parindent}%
     \^^\{\abovedisplayskip}%
     \^^\{\belowdisplayskip}%
     \^^\{\abovedisplayshortskip}%
     \^^\{\belowdisplayshortskip}%
     \^^\{\abovechapterskip}%
     \^^\{\belowchapterskip}%
     \^^\{\abovesectionskip}%
     \^^\{\belowsectionskip}%
     \^^\{\abovesubsectionskip}%
     \^^\{\belowsubsectionskip}}


\def\twoupsetup{
     \topmargin=.75in
     \leftmargin=.5in
     \vsize=6.9in
     \hsize=4.75in
     \fullhsize=10in
     \let\draft=\relax}

\outputstyle{normal}                             

\def\marginnoteformat{\subscriptsize             
     \hsize=1in \baselinestretch=1000 \everypar={}%
     \tolerance=5000 \hbadness=5000 \parskip=0pt \parindent=0pt
     \leftskip=0pt \rightskip=0pt \raggedright}

\head={\ifdraft\normalfonts\it\hfil DRAFT\hfil   
     \llap{\number\day\ \monthword\month\ \militarytime}\else\hfil\fi}
\foot={\hfil\normalfonts\numstyle\pagenum\hfil}  

\normalbaselineskip=12pt                         
\normallineskip=0pt                              
\normallineskiplimit=0pt                         
\normalbaselines                                 

\topskip=.85\baselineskip \splittopskip=\topskip
\headskip=2\baselineskip \footskip=\headskip

\pagenumstyle{arabic}                            

\parskip=0pt                                     
\parindent=20pt                                  

\baselinestretch=1000                            


\chapterstyle{left}                              
\chapternumstyle{blank}                          
\def\chapterbreak{\newpage}                      
\abovechapterskip=0pt                            
\belowchapterskip=1.5\baselineskip               
     plus.38\baselineskip minus.38\baselineskip
\def\chapternumformat{\numstyle\chapternum.}     

\sectionstyle{left}                              
\sectionnumstyle{blank}                          
\def\sectionbreak{\vskip0pt plus4\baselineskip\penalty-100
     \vskip0pt plus-4\baselineskip}              
\abovesectionskip=1.5\baselineskip               
     plus.38\baselineskip minus.38\baselineskip
\belowsectionskip=\the\baselineskip              
     plus.25\baselineskip minus.25\baselineskip
\def\sectionnumformat{
     \ifblank\chapternumstyle\then\else\numstyle\chapternum.\fi
     \numstyle\sectionnum.}

\subsectionstyle{left}                           
\subsectionnumstyle{blank}                       
\def\subsectionbreak{\vskip0pt plus4\baselineskip\penalty-100
     \vskip0pt plus-4\baselineskip}              
\abovesubsectionskip=\the\baselineskip           
     plus.25\baselineskip minus.25\baselineskip
\belowsubsectionskip=.75\baselineskip            
     plus.19\baselineskip minus.19\baselineskip
\def\subsectionnumformat{
     \ifblank\chapternumstyle\then\else\numstyle\chapternum.\fi
     \ifblank\sectionnumstyle\then\else\numstyle\sectionnum.\fi
     \numstyle\subsectionnum.}


\footnotenumstyle{symbols}                       
\footnoteskip=0pt                                
\def\footnotenumformat{\numstyle\footnotenum}    
\def\footnoteformat{\footnotesize                
     \everypar={}\parskip=0pt \parfillskip=0pt plus1fil
     \leftskip=1em \rightskip=0pt
     \spaceskip=0pt \xspaceskip=0pt
     \def\\{\ifhmode\ifnum\lastpenalty=-10000
          \else\hfil\penalty-10000 \fi\fi\ignorespaces}}


\def\undefinedlabelformat{$\bullet$}             


\equationnumstyle{arabic}                        
\subequationnumstyle{blank}                      
\figurenumstyle{arabic}                          
\subfigurenumstyle{blank}                        
\tablenumstyle{arabic}                           
\subtablenumstyle{blank}                         

\eqnseriesstyle{alphabetic}                      
\figseriesstyle{alphabetic}                      
\tblseriesstyle{alphabetic}                      

\def\puteqnformat{\hbox{
     \ifblank\chapternumstyle\then\else\numstyle\chapternum.\fi
     \ifblank\sectionnumstyle\then\else\numstyle\sectionnum.\fi
     \ifblank\subsectionnumstyle\then\else\numstyle\subsectionnum.\fi
     \numstyle\equationnum
     \numstyle\subequationnum}}
\def\putfigformat{\hbox{
     \ifblank\chapternumstyle\then\else\numstyle\chapternum.\fi
     \ifblank\sectionnumstyle\then\else\numstyle\sectionnum.\fi
     \ifblank\subsectionnumstyle\then\else\numstyle\subsectionnum.\fi
     \numstyle\figurenum
     \numstyle\subfigurenum}}
\def\puttblformat{\hbox{
     \ifblank\chapternumstyle\then\else\numstyle\chapternum.\fi
     \ifblank\sectionnumstyle\then\else\numstyle\sectionnum.\fi
     \ifblank\subsectionnumstyle\then\else\numstyle\subsectionnum.\fi
     \numstyle\tablenum
     \numstyle\subtablenum}}


\referencestyle{sequential}                      
\referencenumstyle{arabic}                       
\def\putrefformat{\numstyle\referencenum}        
\def\referencenumformat{\numstyle\referencenum.} 
\def\putreferenceformat{
     \everypar={\hangindent=1em \hangafter=1 }%
     \def\\{\hfil\break\null\hskip-1em \ignorespaces}%
     \leftskip=\refnumindent\parindent=0pt \interlinepenalty=1000 }


\normalsize


\def\fmtversion{2.6M (June 1992)}

\catcode`\@=12

\typesize=10pt \magnification=1200 \baselineskip17truept
\footnotenumstyle{arabic} \hsize=6truein\vsize=8.5truein
\sectionnumstyle{blank}
\chapternumstyle{blank}
\chapternum=1
\sectionnum=1
\pagenum=0

\def\begintitle{\pagenumstyle{blank}\parindent=0pt
\begin{narrow}[0.4in]}
\def\endtitle{\end{narrow}\newpage\pagenumstyle{arabic}}


\def\beginexercise{\vskip 20truept\parindent=0pt\begin{narrow}[10
truept]}
\def\endexercise{\vskip 10truept\end{narrow}}


\def\eql#1{\eqno\eqnlabel{#1}}
\def\ref{\reference}
\def\peq{\puteqn}
\def\pref{\putref}

\def\mgn{\marginnote}
\def\bex{\begin{exercise}}
\def\eex{\end{exercise}}


\font\open=msbm10 


\def\StretchRtArr#1{{\count255=0\loop\relbar\joinrel\advance\count255 by1
\ifnum\count255<#1\repeat\rightarrow}}
\def\StretchLtArr#1{\,{\leftarrow\!\!\count255=0\loop\relbar
\joinrel\advance\count255 by1\ifnum\count255<#1\repeat}}

\def\StretchLRtArr#1{\,{\leftarrow\!\!\count255=0\loop\relbar\joinrel\advance
\count255 by1\ifnum\count255<#1\repeat\rightarrow\,\,}}

\def\mbox#1{{\leavevmode\hbox{#1}}}

\def\hspace#1{{\phantom{\mbox#1}}}
\def\oR{\mbox{\open\char82}}

\def\al{\alpha}
\def\bom{{\bmit\omega}}
\def\be{\beta}
\def\ga{\gamma}
\def\de{\delta}
\def\Ga{\Gamma}

\def\la{\lambda}

\def\om{\omega}

\def\si{\sigma}

\def\th{\theta}

\def\ze{\zeta}

\def\De{\Delta}

\def\caD{{\cal D}}

\def\det{{\rm det\,}}

\def\sc{{\rm sc }}

\def\zf{$\zeta$--function}
\def\zfs{$\zeta$--functions}
\def\hk{heat-kernel}


\def\frac#1/#2{\leavevmode\kern.1em
\raise.5ex\hbox{\the\scriptfont0 #1}\kern-.1em/\kern-.15em
\lower.25ex\hbox{\the\scriptfont0 #2}}
\def\sfrac#1/#2{\leavevmode\kern.1em
\raise.5ex\hbox{\the\scriptscriptfont0 #1}\kern-.1em/\kern-.15em
\lower.25ex\hbox{\the\scriptscriptfont0 #2}}

\def\gtorder{\mathrel{\raise.3ex\hbox{$>$}\mkern-14mu
             \lower0.6ex\hbox{$\sim$}}}
\def\ltorder{\mathrel{\raise.3ex\hbox{$<$}\mkern-14mu
             \lower0.6ex\hbox{$\sim$}}}

\def\semidirprod{\rlap{\ss C}\raise1pt\hbox{$\mkern.75mu\times$}}
\def\for{\lower6pt\hbox{$\Big|$}}
\def\fish{\kern-.25em{\phantom{abcde}\over \phantom{abcde}}\kern-.25em}


\def\boxit#1{\vbox{\hrule\hbox{\vrule\kern3pt
        \vbox{\kern3pt#1\kern3pt}\kern3pt\vrule}\hrule}}
\def\dalemb#1#2{{\vbox{\hrule height .#2pt
        \hbox{\vrule width.#2pt height#1pt \kern#1pt \vrule
                width.#2pt} \hrule height.#2pt}}}

\def\ol{\overline}
\def\frac#1#2{{{#1}\over{#2}}}

\def\comb#1#2{{\left(#1\atop#2\right)}}

\def\cosec{{\rm cosec\,}}

\def\eg{{\it e.g.}}
\def\ie{{\it i.e. }}
\def\cf{{\it cf }}
\def\pa{\partial}

\def\av#1{\langle#1\rangle} 
\def\me#1#2#3{\langle{#1}\mid\!{#2}\!\mid{#3}\rangle}  
\def\rme#1#2#3{\langle{#1}\mid\mid\!{#2}\!\mid\mid{#3}\rangle}  

\def\Tr{{\rm Tr\,}}

\def\curl{{\rm curl\,}}

\def\sumdash#1{{\mathop{{\sum}'}_{#1}}}
\def\sumdasht#1#2{{\mathop{{\sum}'}_{#1}^{#2}}}

\def\Threej#1#2#3#4#5#6{\biggl({#1\atop#4}{#2\atop#5}{#3\atop#6}\biggr)}

\def\3j#1#2#3#4#5#6{\left\lgroup\matrix{#1&#2&#3\cr#4&#5&#6\cr}
\right\rgroup}

\def\man{{\cal M}}

\def\m?{\mgn{?}}

\def\pa{\partial}

\def\beq{\begin{eqnarray}}
\def\eeq{\end{eqnarray}}


\def\cmp#1#2#3{{\it Comm. Math. Phys.} {\bf {#1}} ({#2}) #3}
\def\cqg#1#2#3{{\it Class. Quant. Grav.} {\bf {#1}} ({#2}) #3}

\def\jmp#1#2#3{{\it J. Math. Phys.} {\bf {#1}} ({#2}) #3}
\def\jpa#1#2#3{{\it J. Phys.} {\bf A{#1}} ({#2}) #3}

\def\np#1#2#3{{\it Nucl. Phys.} {\bf B{#1}} ({#2}) #3}
\def\pl#1#2#3{{\it Phys. Lett.} {\bf {#1}} ({#2}) #3}

\def\prp#1#2#3{{\it Phys. Rep.} {\bf {#1}} ({#2}) #3}
\def\pr#1#2#3{{\it Phys. Rev.} {\bf {#1}} ({#2}) #3}

\def\prD#1#2#3{{\it Phys. Rev.} {\bf D{#1}} ({#2}) #3}

\def\rmp#1#2#3{{\it Rev. Mod. Phys.} {\bf {#1}} ({#2}) #3}

\def\zfp#1#2#3{{\it Z. f. Phys.} {\bf {#1}} ({#2}) #3}

\def\cras#1#2#3{{\it Comptes Rend. Acad. Sci. (Paris)} {\bf{#1}} (#2) #3}
\def\prs#1#2#3{{\it Proc. Roy. Soc.} {\bf A{#1}} ({#2}) #3}

\def\amsh#1#2#3{{\it Abh. Math. Sem. Ham.} {\bf {#1}} ({#2}) #3}
\def\am#1#2#3{{\it Acta Mathematica} {\bf {#1}} ({#2}) #3}
\def\aim#1#2#3{{\it Adv. in Math.} {\bf {#1}} ({#2}) #3}
\def\ajm#1#2#3{{\it Am. J. Math.} {\bf {#1}} ({#2}) #3}

\def\aom#1#2#3{{\it Ann. of Math.} {\bf {#1}} ({#2}) #3}
\def\cjm#1#2#3{{\it Can. J. Math.} {\bf {#1}} ({#2}) #3}

\def\dmj#1#2#3{{\it Duke Math. J.} {\bf {#1}} ({#2}) #3}
\def\invm#1#2#3{{\it Invent. Math.} {\bf {#1}} ({#2}) #3}

\def\jdg#1#2#3{{\it J. Diff. Geom.} {\bf {#1}} ({#2}) #3}

\def\jram#1#2#3{{\it J. f. reine u. Angew. Math.} {\bf {#1}} ({#2}) #3}
\def\jims#1#2#3{{\it J. Indian. Math. Soc.} {\bf {#1}} ({#2}) #3}
\def\jlms#1#2#3{{\it J. Lond. Math. Soc.} {\bf {#1}} ({#2}) #3}

\def\ma#1#2#3{{\it Math. Ann.} {\bf {#1}} ({#2}) #3}

\def\mz#1#2#3{{\it Math. Zeit.} {\bf {#1}} ({#2}) #3}
\def\ojm#1#2#3{{\it Osaka J.Math.} {\bf {#1}} ({#2}) #3}

\def\plms#1#2#3{{\it Proc. Lond. Math. Soc.} {\bf {#1}} ({#2}) #3}
\def\pgma#1#2#3{{\it Proc. Glasgow Math. Ass.} {\bf {#1}} ({#2}) #3}
\def\qjm#1#2#3{{\it Quart. J. Math.} {\bf {#1}} ({#2}) #3}

\def\rmjm#1#2#3{{\it Rocky Mountain J. Math.} {\bf {#1}} ({#2}) #3}

\def\tams#1#2#3{{\it Trans.Am.Math.Soc.} {\bf {#1}} ({#2}) #3}

\begin{title}
\vglue 1truein
\vskip15truept
\centertext {\Bigfonts \bf Spherical Universe topology}
\vskip5truept
\centertext {\Bigfonts \bf and the Casimir effect}
\vskip 20truept
\centertext{J.S.Dowker\footnote{dowker@a35.ph.man.ac.uk}} \vskip
7truept \centertext{\it Department of Theoretical Physics,}
\centertext{\it The University of
 Manchester,} \centertext{\it Manchester, England}
\vskip40truept
\begin{narrow}
Recent interest in the possible non--trivial topology of the
Universe, and the resulting analysis of the Laplacian eigenproblem,
has prompted a reprise of calculations done by ourselves some time
ago. The mode problem on the fixed--point--free factored 3--sphere,
S$^3/\Ga$, is re--addressed and applied to some field theory
calculations for massless fields of spin 0, 1/2 and 1. In particular
the degeneracies on the factors, including lens spaces, are rederived
more neatly in a geometric fashion. Likewise, the vacuum energies are
re-evaluated by an improved technique and expressed in terms of the
polyhedrally invariant polynomial degrees, being thus valid for all
cases without angle substitution. An alternative, but equivalent
expression is given employing the cyclic decomposition of $\Ga$. The
scalar functional determinants are also determined. As a bonus, the
spectral asymmetry function, $\eta(s)$ is treated by the same
approach and explicit forms are given for $\eta(-2n)$ on one--sided
lens spaces.

\end{narrow}
\vskip 5truept
\vskip 60truept
\vfil
\end{title}
\pagenum=0
\newpage

\section{\bf 1. Introduction.}

The current interest in  the topology of the Universe has led to
calculations involving the modes on discrete factors of the sphere,
$S^d/\Ga$, with $\Ga$ freely acting. These are required both for the
spectral analysis of the appearance of the Universe and for quantum
field theory calculations.

Weyl raised the question of the topology (`inter--connection') of the
Universe in his classic book `Space--Time--Matter' in 1922 and later
cosmic speculations were made by Ellis [\pref{Ellis}] in connection
with the Friedman--Robertson--Walker metric. Milnor [\pref{Milnor}]
has also considered the observational consequences of a non-trivial
topology. Some other references can be found in [\pref{LaandL,LLL}],
for example. Starkman, [\pref{Starkman}], includes a translation of
the pre-GR work of Schwarzschild, [\pref{Schwarzschild}].

The enumeration of manifolds locally isometric to the sphere, in
particular to the three--sphere, is a textbook matter and factored
spheres occur frequently in various contexts seeing that they provide
examples of multiply connected spaces that are relatively easy to
control. I might mention the topic of analytic torsion. Spheres occur
as hypersurfaces and boundaries and these can be replaced by factored
spheres as in the analysis of boundary terms in the index theorem,
\eg\ Gibbons, Pope and R\"omer, [\pref{GPR}], and in the generalised
cone, [\pref{Cheeger}], [\pref{Dow6}].

Our interest in such spaces was originally as examples in connection
with quantum mechanics on multiply connected spaces. It was suggested
in [\pref{Dow}], for example, that the target space in the
$\si$--model could just as well be $S^3/\Ga$ as $S^3$. Pion
perturbation theory would not distinguish between these. The quantum
mechanical and field theoretic propagators on $S^3/\Ga$ or $T\times
S^3/\Ga$ are given as pre-image sums of those for the full $S^3$. The
nice review by Camporesi [\pref{Camporesi}] contains extensive
information on these sphere, and other homogeneous space, quantities.

In [\pref{DandB}], we presented some field theory calculations on
multiply connected Clifford -- Klein spaces, including the flat
(Hantsche and Wendt), $T\times \oR^3/\Ga$, ones and the curved
(Seifert and Threlfall) ones, $T\times S^3/\Ga$. For simplicity, we
chose those $\Ga$ that produce homogeneous manifolds and the
techniques used involved \zfs\ and images. In later calculations,
concerned with symmetry breaking by `Wilson lines',
[\pref{DandJ,Jadhav2}], similar ingredients were employed.

In the course of the evaluations we naturally encountered mode
properties and expressions for degeneracies. These have occurred in
some recent works, \eg\ [\pref{LWUGL}], dealing with cosmic topology.
In the present work I wish to re--examine these technical questions
while filling in some gaps and extending the earlier discussions. In
[\pref{DandB}], the Casimir energies on lens spaces $S^3/Z_q$ were
given generally in terms of polynomials in $q$ and I wish here to
extend these to prism spaces, $S^3/D'_q$. Some results along these
lines have already been given in [\pref{DandJ}] taken from
[\pref{Jadhav}]. One might also wish to consider the case of
non-homogeneous manifolds, which were only mentioned in
[\pref{DandB}]. Of course, any `realistic' cosmology must be
time--dependent but since my aim here is simply to exhibit some
mathematical details I consider only the static Einstein Universe,
T$\times\man$.

In addition to lens and prism spaces, in [\pref{DandB}] we also
computed the Casimir energies for the other binary polyhedral groups.
An objective is to rederive these, the point being that they turned
out to be rational quantities arising from combinations of terms
containing irrational quantities. The geometric reason is clear.
Roughly, each group can be expressed in terms of cyclic groups and we
then only have to  combine appropriately the above mentioned
polynomials. An elaboration of this might be a good starting point,
however some necessary preliminaries have to be recounted. I refer to
the mode problem.

The expressions for scalar modes on the full $d$--dimensional sphere
go back as far as Green in 1837 and were developed by Hill in 1883,
[\pref{Hill}]. Later discussions naturally abound and have entered
the standard reference works so it is unnecessary to give any sort of
comprehensive history here. In the following section I present some
basic facts.

\section{\bf 2. Modes and degeneracies
 on the three-sphere and factored three-sphere.}

The three--sphere case is special because of the isomorphism SO(4)
$\sim$ SU(2)$\times$ SU(2)$/Z_2$, essentially a consequence of the
isomorphism $S^3\sim$ SU(2), the two factors corresponding to left
and right group actions. The fact that, for a free, discrete action,
the factors must be binary polyhedral groups was derived by Seifert
and Threlfall, [\pref{SeandT}], although known to Hopf,
[\pref{Hopf}]. The classic discussions of these groups are due to
Klein, [\pref{Klein}],  and Cayley, [\pref{Cayley}]. There are, of
course, many later treatments. Wolf, [\pref{Wolf}], is one standard
reference and he also treats the $d$--sphere, see also Milnor,
[\pref{Milnor2}]. Handy information is available in Coxeter and
Moser, [\pref{CandM}], and in Coxeter, [\pref{Coxeter2}].

The binary groups also occur in the quantum mechanics of electrons in
crystals, the original examination being by Bethe, [\pref{Bethe}]. He
calls them {\it double groups} and his technique has passed into
physics textbooks, \eg\ Landau and Lifshitz, [\pref{LandL}]. A more
rigorous analysis is provided by Opechowski, [\pref{Opechowski}]. The
standard mathematical reference, [\pref{Coxeter2}], does not mention
Bethe's work.

The spatial manifold I am concerned with here is, therefore,
$\man=$S$^3/(\Ga_L\times\Ga_R)$ with, to repeat, both $\Ga_L$ and
$\Ga_R$ binary polyhedral groups. For a homogeneous space, one of the
factors will be trivial, equal to ${\bf 1}$, but for a while I keep
to the general situation.

One final scene--setting point has to be raised before the
calculation is begun. The quantum mechanics, and therefore scalar
quantum field theory, on spaces with a non--trivial first homotopy
group, $\pi_1(\man)$, (which is isomorphic to $\Ga$ for free actions)
has a freedom coded by the homomorphism, $\pi_1(\man)\to$ U(1). I do
not wish to invoke this freedom in the following. It can easily be
incorporated but to do so would extend the algebra, and this paper,
unnecessarily.

In order to evaluate the Casimir energy, for example, one needs the
equations of motion on T$\times\man$. This amounts to a choice of
scalar Laplacian on $\man$. One choice is the bare Laplacian, $\De$,
and another is the `conformal' Laplacian $\De+R/6$, on S$^3$. (We
define $\De$ with the sign such that its spectrum is non--negative.)
This choice will affect the eigenvalues but not the degeneracies nor
the eigenfunctions, and, since it is these I wish to spotlight, I
work, for preference, with the conformal Laplacian which makes the
eigenvalues on the full sphere squares of integers, say $l^2/a^2$,
$l=1,2,\ldots$, up to a scaling, $a$ being the radius. The Laplacian
on S$^3$ coincides with the Casimir operator on SU(2), up to a scale,
and the eigenfunctions can be taken as proportional to the complete
set of representation matrices, $\caD^j_{mn}(g)$, $g\in$ SU(2) with
dimensions, $l=2j+1$. This is true for any Lie group. Square
integrable completeness is the content of the Peter--Weyl theorem.
The classic book by Vilenkin, [\pref{Vilenkin}], provides all the
details one could require. Talman, [\pref{Talman}], and Miller,
[\pref{Miller}], are also very useful. In this paper I restrict
attention to the three--sphere where one has the full array of
angular momentum techniques to play with.

As is well known, going back at least to Rayleigh, the effect of the
factoring, $\man\to\man/\Ga$, amounts to a cull of the modes on
$\man$. In solid state physics this process is referred to as
symmetry adaptation and functions on $\man/\Ga$ can be obtained by
projection from those on $\man$, which amounts to averaging over
$\Ga$. This process can be traced back, in its general form, to
Cartan and Weyl. Making this projection does not always immediately
yield quantities of practical value.

Stiefel, [\pref{Stiefel}], makes some useful remarks on the
application of group theory to the solution of boundary value
problems.

One must begin therefore, again, with the scalar modes on the full
sphere, S$^3$, for which it is sufficient to take the hyperspherical
harmonics, $\caD^j_{mn}(g)$.

If one requires the explicit form of the eigenmodes, then the
traditional method is to select an appropriate coordinate system,
separate variables and solve some ordinary differential equations, by
various means. In this way, the modes were known to Green for
arbitrary dimensions, were developed by Hill and related to ambient
harmonic polynomials. This is the way the $\caD^j_{mn}(g)$ are
usually evaluated in standard angular momentum references using, for
example, Euler angles and involving Jacobi polynomials. Vilenkin,
[\pref{Vilenkin}], has the details, and much else.

As a rule, it is more elegant to use as much group theory (here
angular momentum theory) as possible.

Instead of the $\caD^j_{mn}(g)$ an equivalent set of (scalar)
harmonics may be defined by some left-right recoupling,
  $$
  Y_{n:LM}(g)=\bigg[{(2J+1)(2L+1)\over|\man|}\bigg]^{1/2}
  \Threej JLm{m'}MJ {\caD^J_m}^{m'}(g)\,,
  $$
where $n=2J+1$ and $|\man|$ is the volume of SU(2), $2\pi a^3$.

For some purposes these functions are more convenient than the
$\caD$'s. They are associated with the polar coordinate system ,
$(\chi,\xi,\eta)$ on S$^3\sim$ SU(2). A group element, $g\in$ SU(2),
is parametrised by an angle of rotation, $2\chi$, and the S$^2$
angles, $\xi,\eta$, specify an axis of rotation, using the language
of rotation in the light of the isomorphism, SO(3) $\sim$
SU(2)$/Z_2$.

Explicit formulae for the $Y_{n:LM}$ are derived in the literature
(\eg\ Talman, [\pref{Talman}], Bander and Itzykson, [\pref{BandI}]).
There is a neater method than the one used in these references but,
since the eigenfunctions are not needed in this paper, I leave it
until a later time. Formally I just use the $\caD$'s.

\begin{ignore}

but we would like to take this opportunity to indicate an alternative
derivation. Using the orthogonality  of the $3j$ symbols we wish to
evaluate
  $$
  (-1)^J\bigg[{(2J+1)(2L+1)\over|\man|}\bigg]^{-1/2}
  Y_{n:LM}(g)=\Tr\big(u_M^L(J)\caD^J(g)\big)\,,
  \eql{trace1}
  $$
where
 $$
 \big[u_M^L(J)\big]_m^{\cdot m'}=\Threej JL{m'}mMJ\,.
 $$

The matrices $u^L_M(J)$ are discussed by Racah, [\pref{Racah1}], and
form a complete set in terms of which any $(2J+1)$--square matrix can
be expanded. They have the useful property that, considered as rank
$L$ tensor operators,
  $$
  \rme J{u^L}J=1\,,
  $$
where the reduced matrix element has been defined by the
Wigner--Eckart theorem, (\eg\ Fano and Racah, [\pref{FandR}],
(14.4)),
  $$
  \me {jm}{T_K^q}{j'm'}=\Threej jq{m'}mK{j'}\rme j{T_K}{j'}\,.
  $$

Angular momentum theory is beset by different conventions, mostly of
signs, and it will be as well here to state that ours are generally
those of Fano and Racah, except that our $3j$ symbols are the more
usual ones of Wigner and that our Euler angle and rotation
conventions follow those of Brink and Satchler, [\pref{BandS}].

Since comparison of all the different conventions is a painful
process, we give a few linking remarks. (Some useful comments will be
found in Edmonds, [\pref{Edmonds}], and in Brink and Satchler.) To go
from Fano and Racah's $\caD$'s to those of Brink and Satchler, make
the replacements $(\psi,\th,\phi)\to(-\al,-\be,-\ga)$. Further Brink
and Satchler's $\caD$'s are obtained from those of Talman by the
replacements $(\al,\be,\ga)\to(\al+\pi/2,\be,\ga-\pi/2)$

The method of evaluating the trace in (\peq{trace1}) will now be
outlined. The idea is to replace the $u^L_M$ matrix operator by an
equivalent differential operator which can then be taken outside the
trace and so acts on $\Tr\caD(g)$, which is just the character,
$\chi^J(g)$.

The differential operator is constructed using the basic group
property
  $$
   X_m\caD^J(g)=iJ_m\caD^J(g)\,,
  $$
which displays the replacement of a rank one matrix operator, $i{\bf
J}$, the spin-$j$ angular momentum operator, by a differential one,
${\bf X}$, acting on a representation. The $X_m$ are the left
infinitesimal operators (generators) of the group and are known
explicitly in terms of the derivatives with respect to the
coordinates.

Higher rank operators are constructed from products of angular
momentum matrices. The irreducible products are spherical harmonics
with ${\bf J}$ as argument. They correspond to the `operator
spherical harmonics' of Schwinger, [\pref{Schwinger}], and can be
defined by recursion starting from ${\bf J}$, or by polarisation of
ordinary spherical harmonics (\eg\ Brink and Satchler
[\pref{BandS}]). They are given in this form by Rose [\pref{Rose}].
\end{ignore}

The projected eigenfunctions on S$^3/\Ga$ are periodised sums on
S$^3$ in the standard way,
  $$
  \phi^j_{mn}(g)=\bigg[{2j+1\over2\pi^2 a^3|\Ga_L||\Ga_R|}
  \bigg]^{1/2}
  \sum_{\ga=(\ga_L,\ga_R)} \caD^j_{mn}(\ga_Lg\ga_R)\,.
  \eql{persum1}
  $$

This is an example of the more formal, and general, statement that,
if $\widetilde\phi_{\la_n}(q)$ are the eigenfunctions on the covering
manifold, $\widetilde\man$, then, [\pref{DandB,BandD}],
  $$
  \phi_{\la_n}(q)={1\over\sqrt{|\Ga|}}\sum_\ga
  \widetilde\phi_{\la_n}(\ga q)\,,\quad q\in\man\,,
  \eql{proj1}
  $$
are periodic eigenfunctions on $\man=\widetilde\man/\Ga$. For
convenience, I make no notational distinction between points, $q$, of
$\widetilde\man$ and $\man$.

The standard difficulty is that these projected eigenfunctions are
not independent, as constructed, and a certain amount of
diagonalisation is required. I summarise this well known state of
affairs in the present notation.

From general self--adjointness arguments, both $\phi_{\la_n}(q)$ and
$\widetilde\phi_{\la_n}(q)$ must be orthogonal, on $\man$ and
$\widetilde\man$ respectively, for distinct eigenvalues. They will
also form complete sets. Orthogonality means that one can work
eigenspace by eigenspace.

Label, in the usual way, the covering eigenfunctions by the
eigenvalue $\la$ and an index $i$ to take care of any degeneracy.
Instead of (\peq{proj1}) then, define
   $$
  \phi_{\la,i}(q)={1\over\sqrt{|\Ga|}}\sum_\ga
  \widetilde\phi_{\la,i}(\ga q)\,,\quad q\in\man\,,
  \eql{proj2}
  $$
and construct the scalar product
  $$
  P_{ij}\equiv\int_{\man}dq\,\phi^*_{\la,i}(q)\phi_{\la,j}(q)\,.
  \eql{sp1}
  $$

Using completeness and eigenspace orthogonality, it is easy to show
that $P$ is a projection operator, $P^2=P$. For the proof start with,
  $$
  P_{ij}P_{jk}=\sum_j\int_{\man}dq\int_{\man}dq'\,
  \phi^*_{\la,i}(q)\phi_{\la,j}(q)
  \phi^*_{\la,j}(q')\phi_{\la,k}(q')\,,
  \eql{p2}
  $$
and consider the quantity,
  $$
  \sum_j\phi_{\la,j}(q)  \phi^*_{\la,j}(q')\,.
  \eql{pc1}
  $$
One has completeness on $\man$,
  $$
  \sum_\la\sum_j\phi_{\la,j}(q)
  \phi^*_{\la,j}(q')=\de(q,q')=\sum_\ga\widetilde\de(\ga q,q')\,.
  \eql{comp1}
  $$

Incidentally, this is consistent with the factors in (\peq{proj2})
after a group translation. The left--hand side of (\peq{comp1}) is
  $$\eqalign{
  {1\over|\Ga|}\sum_{\ga,\ga'}\sum_{\la,j}
  \widetilde\phi_{\la,j}(\ga q)
  \,\widetilde\phi^*_{\la,j}(\ga' q')&=
  {1\over|\Ga|}\sum_{\ga,\ga'}\widetilde\de(\ga q,\ga' q')\cr
  &=
  {1\over|\Ga|}\sum_{\ga,\ga'}\widetilde\de(\ga'^{-1}\ga q,q')\cr
  &=\sum_{\ga}\widetilde\de(\ga q,q')\,.
  }
  \eql{comp2}
  $$

Using eigenvalue--$\la$ orthogonality on $\man$, the quantity
(\peq{pc1}) occurring in (\peq{p2}) can be replaced by the full
quantity (\peq{comp1}) and the double integral reduced to a single
one recognised as $P_{ik}$ as required.

Orthogonality on $\man$ implies the following identity on the
covering space
  $$
  {1\over|\Ga|}\sum_\ga\int_{\widetilde\man}
  \widetilde\phi^*_{\la,i}(\ga q)\,\widetilde
  \phi_{\la',j}(q)\,dq=\de_{\la\la'}P_{ij}\,,
  \eql{ciden}
  $$
and completeness on $\widetilde\man$, used in (\peq{comp2}), is
  $$
   \sum_{\la,i}\widetilde\phi_{\la,i}(q)
  \,\widetilde\phi^*_{\la,i}(q')=\widetilde\de(q,q')\,.
  \eql{comp3}
  $$

Now consider the combination
  $$
  \sum_i\widetilde
  \phi_{\la,i}(q)\,P_{ij}={1\over|\Ga|}\sum_\ga\int_{\widetilde\man}dq'\,
  \sum_i\widetilde\phi_{\la,i}(q)\,\widetilde\phi^*_{\la,i}(\ga q')\,
  \widetilde\phi_{\la,j}(q')
  \eql{comb}
  $$
using either (\peq{sp1})+(\peq{proj2}) or (\peq{ciden}). Replace the
sum over $i$ in (\peq{comb}) by the complete sum (\peq{comp3}) and
use (\peq{ciden}) to show that the sum over $\la$ is restricted to
the single term $\la=\la$ and so makes no change, but the integral
can now be performed and I regain the sum (projection) in
(\peq{proj2}), so that
  $$
  \phi_{\la,j}=\sqrt{|\Ga|}\sum_i\widetilde\phi_{\la,i}(q)\,P_{ij}
  $$
which is the algebraic expression of the projection
$\widetilde\man\to\man$.

The diagonalisation referred to earlier is more precisely that of
$P$, which has eigenvalues $1$ and $0$, the number of 1's, \ie $\Tr
P$, being just the degeneracy of the $\la$ level on $\man$. There is
no need to perform the diagonalisation to determine this.

Therefore the degeneracy of the $\la$ eigenvalue is
  $$
  d_\la={1\over|\Ga|}\sum_\ga\int_{\widetilde\man}\sum_i
  \widetilde\phi^*_{\la,i}(\ga^{-1} q)\,\widetilde
  \phi_{\la,i}(q)\,dq\,.
  \eql{degen1}
  $$

Diagonalisation would be required to determine the independent modes
in this direct approach which is not necessarily a practical one.

Equation (\peq{degen1}) is a standard result in the theory of
symmetry adaptation, familiar in quantum mechanics and applying it to
(\peq{persum1}) yields, after some mild group theory, [\pref{DandB}],
  $$
  d_l={1\over|\Ga_L||\Ga_R|}\sum_{\ga=(\ga_l,\ga_R)}\,\chi_l(\ga_L)
  \chi_l(\ga_R)\,,
  \eql{degen2}
  $$
where $\chi_l(g)$ is the character of the spin--$j$ representation,
with $l=2j+1$,
  $$
  \chi_l(g)={\sin l\th\over\sin\th}\,.
  $$
$a\th$ is the radial distance on S$^3$ between the origin,
corresponding to the unit element of SU(2), and the point $q$,
corresponding to the group element, $g$. The character is a class
function. $\th$ is the colatitude in the polar coordinate system on
SU(2). It was denoted by $\chi$ earlier and $2\th=\om$ equals, as
mentioned, the SO(3) rotation angle.

One thus encounters in (\peq{degen2}) the quantities
  $$
  d_l(\Ga)={1\over|\Ga|}\sum_{\ga\in\Ga}\chi_l(\th_\ga)=
  {1\over|\Ga|}\sum_{\ga\in\Ga}{\sin l\th_\ga\over\sin\th_\ga}\,,
  \eql{degen3}
  $$
which can be evaluated for each binary polyhedral group, if desired,
since the angles, $\th_\ga$, are known and the conjugacy class
decompositions can be used to ease the arithmetic. An example for the
ordinary cubic group, ${\bf O}$, is given by Stiefel,
[\pref{Stiefel}].

Of course, the degeneracy is often combined with other quantities in
an eigenmode sum over $l$ and then it may be advantageous to leave
(\peq{degen3}) alone. For example the conformal \zf\ on S$^3/\Ga$ is
  $$
  \ze_\Ga(s)=a^{2s}\sum_{l=1}^\infty {d_l(\Ga_L)d_l(\Ga_R)\over l^{2s}}\,,
  \eql{czeta}
  $$
and the sum over $l$ produces two Epstein \zfs, in this case. An
expression is given later.

Another example is the generating function for $\chi_l$,
  $$
  \sum_{l=1}^\infty \chi_l(\th)\,e^{-2\ga
  l}={1\over2}{1\over\cosh(2\ga)-\cos\th}\,,
  \eql{genfun1}
  $$
obtained by trivial geometric summation. This can often be used for
{\it ad hoc} evaluations. For example, it directly yields the
standard generating function for lens space degeneracies. For
$\Ga=Z_q$ the angles $\th_\ga$ are $\th_p=p(2\pi/q)$ for
$p=0,1,\ldots,q-1$ and so one is led to the (binary) cyclic
generating function (heat--kernel) setting $t=e^{-2\ga}$,
  $$\eqalign{
  G(t,q)=\sum_{l=1}^\infty d_l(q)t^l&={1\over2q}\sum_{p=0}^{q-1}
  {1\over\cosh(2\ga)-\cos (2p\pi/q)}\cr
  &={t(1+t^q)\over(1-t^2)(1-t^q)}\,.
  }
  \eql{genfun2}
  $$
Expansion of the right--hand side is sufficient to yield expressions
for the cyclic degeneracies. If $q$ is even, it follows that $d_l(q)$
is zero for $l$ even, and for $l$ odd we can use the SO(3) result,
  $$
  g(\si,q)=\sum_{l=0}^\infty(2[l/q]+1)\,\si^l={1\over1-\si}
  {1+\si^q\over1-\si^q}\,,
  \eql{genfun6}
  $$
to read off the degeneracy, $d_{2l+1}(q)$, having relabelled $l\to
2l+1$ and set $\si=t^2$.

It is, nevertheless, still of interest to look at the expression
(\peq{degen3}) directly. Similar finite trigonometric sums have been
considered for many years. Most involve sums related to cyclic
groups, $Z_q$. The basic sum is classic and given in Bromwich,
[\pref{Bromwich}] p.272, Ex.18,
  $$
  \sum_{p=1}^{q-1}{\sin(kl\pi p/q)\over\sin(l\pi p/q)}=q-k\,,\quad
  {\rm for}\,\,(l,q)=1\,,
  $$
and $k$ odd, $k<2q-1$.

Using this formula, one can show that
  $$
  \sum_{p=1}^{q-1}{\sin(2\pi tp/q)\over\sin(2\pi p/q)}
  =\bigg\{\matrix{
  -t&\,t\,\,{\rm even}\cr q-t&\,t\,\,{\rm odd}}\,,
  $$
for all integer $t$ and $q$, $0<t<q$. This allows one to find the
$Z_q$ group--averaged SU(2) character,
  $$\eqalign{
  d_l(q)=\langle\chi_l\rangle_q&={1\over q}\sum_{p=0}^{q-1}
  \chi_l(2\pi p/q)\cr
  &=\bigg\{\matrix{r&\,t\,\,{\rm even}\cr r+1&\,t\,\,{\rm
  odd}}\,\bigg\}\,\,q\,\, {\rm odd}\cr
  &=\bigg\{\matrix{0&\,l\,\,{\rm
  even}\cr 2r+1&\,l\,\,{\rm odd}}\,\bigg\}\,\,q\,\, {\rm even}\cr
  }
  \eql{su2char}
  $$
where I have made the mod $q$ residue class  decomposition, $l=rq+t$,
\ie $r=[l/q]$.

These results are of course in agreement with the preceding
calculations. The last result in (\peq{su2char}), with $q\to2q$ and
$l\to2l+1$, is equivalent to the SO(3) character sum,
  $$
  {1\over q}\sum_{p=0}^{q-1}{\sin\big((2l+1)\pi p/q\big)
  \over\sin(\pi p/q)}=
  2[l/q]+1\,,\quad l=0,1,\ldots\,.
  \eql{so3char}
  $$

These are all standard results and, in particular, (\peq{su2char})
gives the Laplacian degeneracies on simple lens spaces, when
multiplied by the left degeneracy, $d_l({\bf 1})=l$, according to
(\peq{degen2}).

The analysis can be extended to general lens spaces by using linked
two-sided actions so that $\ga$ is labelled by $\th_L$ and $\th_R$ as
follows. Going over to the combinations,
  $$
 \al=\th_R+\th_L\,,\quad\be=\th_R-\th_L\,,
 \eql{angles2}
  $$
the lens space, $L(q;l_1,l_2)$, is defined by setting
  $$
  \al={2\pi p\nu_1\over q}\,,\quad \be={2\pi p\nu_2\over q}\,,
  \eql{angles1}
  $$
where $p,\,=0,1,\ldots,q-1\,,$ labels $\ga$. $\nu_1$ and $\nu_2$ are
integers coprime to $q$, with $l_1$ and $l_2$ their mod $q$ inverses.
The simple, `one-- sided' lens space, $L(q;1,1)$, corresponds to
setting $\nu_2=\nu_1=\nu=1$, say, so that $\th_L=0$ and $\th_R=2\pi
p/q$.

The degeneracy is,
  $$\eqalign{
  d_l(q;l_1,l_2)&={1\over q}\sum_{p=0}^{q-1}{\sin (l(\al-\be)/2)
  \over\sin((\al-\be)/2)}{\sin (l(\al+\be)/2)
  \over\sin((\al+\be)/2)}\cr
  &={1\over q}\sum_{p=0}^{q-1}{\cos l\al-\cos l\be\over
  \cos\al-\cos\be}\,.
  }
  $$

It is convenient to leave off the group average and consider the
(partial) generating function
  $$\eqalign{
  \sum_{l=1}^\infty d_l(\al,\be)t^l&\equiv\sum_{l=0}^\infty
  {\cos l\al-\cos l\be\over\cos\al-\cos\be}\,t^l\cr
  &=t(1-t^2)\bigg({1\over 1+t^2-2t\cos\al}\bigg)
  \bigg({1\over 1+t^2-2t\cos\be}\bigg)\,,
  }
  \eql{genfun3}
  $$
using the elementary summation, \cf\ (\peq{genfun1}),
  $$
  2\sum_{l=0}^\infty \cos (l\al) \,t^l=1+{1-t^2\over1+t^2-2t\cos\al}\,.
  \eql{elemsum1}
  $$

(\peq{genfun3}) is the same generating function derived by Ray
[\pref{Ray}]. (Actually he does $p$--forms and $d$--spheres.)

The full degeneracy follows upon averaging over the group elements,
\ie the angles $\al$ and $\be$ given, for a lens space, by
(\peq{angles1}). Except for the one--sided case, $\al=\pm\be$, it
does not seem possible to complete the sum over $p$. In this
particular case we obtained (\peq{genfun2}) for the one--sided
degeneracy and this can also be found, as a check, from
(\peq{genfun3}) setting $\al=\be$, say, and using the integrated form
of (\peq{genfun3}),
  $$\eqalign{
  \sum_{l=1}^\infty {d_l(\al,\al)\over l}\,t^l
  ={t\over 1+t^2-2t\cos\al}\,.
  }
  \eql{genfun4}
  $$
The division by $l$ on the left corresponds to the removal of the
left degeneracy.

Turning to the other binary groups, we need their structure, which
is, of course well documented.

The ordinary polyhedral groups, considered as subgroups of SO(3),
have a natural action on the two--sphere. They are generated by
rotations through $2\pi/\la,2\pi/\mu,$ $2\pi/\nu$ about the vertices
of a spherical triangle of angles $\pi/\la,\pi/\mu,\pi/\nu$ on S$^2$.
A fundamental domain is comprised of such a triangle together with
its reflection. For the dihedral group, $D_q$, the fundamental domain
can be taken to be the lune, or digon, of apex angle, $\pi/q$.

The binary groups are obtained by lifting the action of the ordinary
ones using the isomorphism, SO(3)$=$SU(2)/Z$_2$. Opechowski,
[\pref{Opechowski}], for example, spells this out.

Coxeter and Moser denote the ordinary group by $(\la,\mu,\nu)$ and
its double by $\langle\la,\mu,\nu\rangle$. The double of an ordinary
group $G$ is variously denoted by $G'$, [\pref{LandL}], $G^{\dag}$,
[\pref{Opechowski}], $2G$, [\pref{Warner}], $G^*$, [\pref{Wolf}].

The lifting can be accommodated geometrically by replacing the
two--sphere by a two-sheeted Riemann surface with branch points at
the vertices of the above spherical triangulation, [\pref{Coxeter2}],
which is, of course, the same triangulation that results from the
application of the complete symmetry groups of the regular solids.

Algebraically, this doubling is mirrored by the formal introduction,
following Bethe, into the presentation of the group of an element,
denoted $Q$, that commutes with the other generators and satisfies
$Q^2=E$, $(E\equiv{\rm id})$. $Q$ corresponds to a rotation through
$2\pi$. The double group $\langle\la,\mu,\nu\rangle$ is generated by
$L,M$ and $N$ with relations
  $$\eqalign{
  &L^\la=M^\mu=N^\nu=LMN=Q\cr
  Q^2&=E\,,\,[L,Q]=[M,Q]=[N,Q]=0\,.
  }
  $$

I first look at the group with an infinite number of members. This is
the binary dihdral group, $D_q'$. Because, for two--sided actions,
one has to substitute in the angles, $\th_\ga$, by hand I consider
only right actions. I choose to write the generator--relation
structure as,
  $$
  A^q=B^2=(AB)^2=Q\,,\quad Q^2=E\,,
  $$
and can thus formally write $D_q'$ as the direct sum
  $$
  D_q'=Z_{2q}\oplus Z_{2q}B\,,
  $$
where $Z_{2q}$ is generated by $A$.

The angles $\th_\ga$ are
  $$
  \th_\ga=\pi p/q\,,\quad\pi\mp\pi p/q\,,\quad p=0,\ldots,q-1
  $$
for $A^p$. The minus sign adjusts the range of $\th$ to be between
$0$ and $\pi$, as is appropriate for the colatitude in polar
coordinates on S$^3$. Equivalently, in order to be more in tune with
the action on the doubly covered two--sphere, $\th$ can be `unrolled'
to run from $0$ to $2\pi$, as on a circle, a great circle in fact.
Doing this corresponds to taking the plus sign. Remember, the angle
$\th$ is half the rotation angle. (Actually $\th$ can be completely
unrolled to be a coordinate on the real line, but this is not
relevant here.)

For $\ga=A^pB$, \ie those $2q$ elements containing a
(binary) dihedral rotation, $\th_\ga=\pi/2$ for all $\ga$.

Hence, from (\peq{degen3}), the right action degeneracy is,
  $$
  d_l(D_q')={1-(-1)^l\over4q}\sum_{p=0}^{q-1}
  {\sin(l\pi p/q)\over\sin(\pi
  p/q)}+{1\over2}\sin(l\pi/2)\,,
  $$
so that $l$ is restricted to odd values, when, with $l\to 2l+1$,
  $$
  d_{2l+1}(D_q')=[l/q]+{1\over2}(1+(-1)^l)\,,\quad l=0,1,\ldots\,,
  \eql{dhdegen}
  $$
using the SO(3) formula (\peq{so3char}).

A generating function can also be found. Having got
(\peq{dhdegen}), a simple way is to use (\peq{genfun6}) which yields
  $$
  \sum_{l=0}^\infty d_{2l+1}(D_q')\si^l={1\over2}\bigg({1\over1-\si}
  {1+\si^q\over1-\si^q}+{1\over1+\si}\bigg)
  ={1+\si^{q+1}\over(1-\si^2)(1-\si^q)}\,.
  \eql{gfdi1}
  $$
Recall that the {\it full} degeneracy on S$^3$ is obtained by
multiplying by the left action degeneracy, $2l+1$, to give,
  $$
  (2l+1)\,d_{2l+1}(D_q')\,.
  $$

Note that the S$^3$ formula for the essential part, (\peq{gfdi1}), of
the right generating function for $D_q'$ actually  coincides with the
S$^2$ formula for $D_q$. Similar considerations hold for the other
double groups as I now discuss.

Referring to the formula for the right degeneracy on S$^3/\Ga'$,
(\peq{degen3}), the doubling means that for every $\th_\ga$ between
$0$ and $\pi$ there is another, $\th_\ga+\pi$, in the range $\pi$ to
$2\pi$. Hence one can write
   $$\eqalign{
  d_l(\Ga')&={1\over|\Ga'|}\sum_{\ga\in\Ga'\atop{0\le\th_\ga<\pi}}
  {\sin l\th_\ga-\sin l(\pi+\th_\ga)\over\sin\th_\ga}\cr
  &={1-(-1)^l\over2|\Ga|}\sum_{\ga\in\Ga'\atop{0\le\th_\ga<\pi}}
  {\sin l\th_\ga\over\sin\th_\ga}\,,
  }
  \eql{degen5}
  $$
whence $l$ is odd $=2l+1$ so
  $$
  d_{2l+1}(\Ga')={1\over|\Ga|}\sum_{\ga\in\Ga}
  {\sin (2l+1)\th_\ga\over\sin\th_\ga}=
  {1\over|\Ga|}\sum_{\ga\in\Ga}
  {\sin (2l+1)\om_\ga/2\over\sin\om_\ga/2}
  \eql{degen6}
  $$
which is the scalar Laplacian degeneracy on the rotational orbifold,
S$^2/\Ga$, denoted $d(l;\Ga)$. This is best discussed as follows.

For the purely rotational polyhedral groups, let $n_q$ be the number
of conjugate $q$--fold axes. Then the S$^2/\Ga$ scalar Laplacian
degeneracy is (\cf\ [\pref{Stiefel,ChandD}]),
  $$\eqalign{
  d(l;\Ga)&=(1-\sum_q n_q){2l+1\over|\Ga|}+{1\over|\Ga|}\sum_q
  qn_qd_q(l)\cr
  &={1\over|\Ga|}\sum_q qn_qd_q(l)-{2l+1\over2}\,,\quad l=0,1,\ldots\,,
  }
  \eql{2spdegen}
  $$
where $d_q(l)$ is the $Z_q$ cyclic degeneracy on S$^2$ given above as
$d_q(l)=2[l/q]+1$. The final equality does not hold for $\Ga$ itself
a cyclic group.

Thus, on S$^2$, all that is necessary is to combine the cyclic
degeneracies, [\pref{Meyer,PandM}]. Expressing things rather in terms
of generating functions, for the {\it two}--sphere we have,
  $$\eqalign{
  g(\si;\Ga)&\equiv\sum_{l=0}^\infty d(l;\Ga)\si^l\cr
  &={1\over|\Ga|}\sum_q qn_qg(\si,q)-{1\over2}g(\si,1)\,,
  }
  \eql{genfunm}
  $$
where $g(\si,q)=g(\si;Z_q)$ is given by (\peq{genfun6}) and
$g(\si,1)= g(\si;{\bf 1})$.

For example, for the dihedral group, $D_q$, $n_q=1$, $n_2=q$ and
simple arithmetic gives,\mgn{meyer2.dfw}
  $$
  g(\si;D_q)={1+\si^{1+q}\over(1-\si^2)(1-\si^q)}\,,
  \eql{gfdi2}
  $$
agreeing with (\peq{gfdi1}), and is the standard formula for the
dihedral Poincare series, \eg\ [\pref{Meyer,BandB}]. The powers of
$\si$ on the denominator are the degrees associated with the
dihedrally invariant polynomial basis, \eg\ [\pref{Springer}].

For the regular solids (not the dihedron), (\peq{2spdegen}) and
(\peq{genfunm}) simplify on application of the orbit--stabiliser
relation, $|\Ga|=2qn_q\,,\forall q$,
  $$
  d(l;\Ga)
  ={1\over2}\bigg(\sum_q d_q(l)-2l-1\bigg)\,,\quad l=0,1,\ldots\,,
  \eql{2spdegen2}
  $$
  and
  $$
  g(\si;\Ga)
  ={1\over2}\bigg(\sum_q g(\si,q)-g(\si,1)\bigg)\,,
  \eql{genfunm2}
  $$
which is a rather neat result.

As an example take the octahedral group ${\bf O}$, for which $n_2=6,
n_3=4$ and $n_4=3$. Simple arithmetic yields
  $$
  g(\si;{\bf O})={1+\si^9\over(1-\si^4)(1-\si^6)}\,,
  \eql{octgenfun}
  $$
for the generating function, obtainable in other ways.

We can use the identity
 $$
 d_{2l+1}(\Ga')=d(l;\Ga)\,,
 \eql{diden}
  $$
together with (\peq{2spdegen2}) to get the right degeneracies on
S$^3/{\bf \Ga'}$, most easily,
  $$\eqalign{
   d_{2l+1}({\bf O}')&=d(l;{\bf O})\cr
   &=[l/2]+[l/3]+[l/4]+1+l\cr
 d_{2l+1}({\bf T}')&=[l/2]+2[l/3]+1+l\cr
 d_{2l+1}({\bf Y}')&=[l/2]+[l/3]+[l/5]+1+l\,.\cr
 }
 \eql{restdegen}
  $$
These results are therefore better appreciated as having an $\oR^3$
geometric origin.

The derivation of these somewhat standard formulae given by Ikeda,
[\pref{Ikeda}], is more involved although he does treat higher
spheres. His technique is algebraic and involves resolving  the
groups into subgroups. His expressions for the lens space
degeneracies differ in form from mine.

As I have remarked, the corresponding evaluations in the case of
double sided actions are much harder. Ikeda and Yamamoto,
[\pref{IandY}], examine two--sided lens spaces.
\section{\bf 3. Heat--kernels and partition functions.}
On the unit three-sphere the eigenvalues of the conformal Laplacian
equal $l^2$, $l=1,\ldots$ and so the integrated heat--kernel
associated with the {\it square--root} of this Laplacian (the
so--called {\it cylinder} kernel) on S$^3/\Ga'$ equals
  $$
  K^{1/2}(\tau)=\sum_{l=1}^\infty ld_l(\Ga')\,e^{-l\tau}\,,
  \eql{rootk}
  $$
which on setting $t=e^{-\tau}$ is recognised as a generating
function. This can be related to the polyhedral generating functions
$g(\si;\Ga)$ as follows. Define
  $$
  G_{\rm tot}(t;\Ga')=K^{1/2}(\tau)
  =\sum_{l=1}^\infty l\,d_l(\Ga')\,t^l\,.
  \eql{rootk2}
  $$

The filtering process giving the eigenproblem on S$^3/\Ga'$ restricts
$l$ to odd values, as has been shown,  and so
  $$\eqalign{
  G_{\rm tot}(t;\Ga')&=\sum_{l=0}^\infty (2l+1)\,d_{2l+1}(\Ga')
  \,e^{-(2l+1)\tau}\cr
  &=-{d\over d\tau}\sum_{l=0}^\infty \,d_{2l+1}(\Ga')\,
  e^{-(2l+1)\tau}\,\cr
  &=-{d\over d\tau}\,e^{-\tau}\,\sum_{l=0}^\infty \,d(l;\Ga)\,e^{-2l\tau}\cr
  &=-{d\over d\tau}\,e^{-\tau}\,g(\si;\Ga)\,,
  }
  \eql{gentot}
  $$
with $\si=t^2=e^{-2\tau}$.
\begin{ignore}
This can be written
  $$
  {1\over t}G_{\rm tot}(t;\Ga')=g(\si;\Ga)+2\si{d\over
  d\si}g(\si;\Ga)\,.
  $$
\end{ignore}

As an organisational point I note that these results do not apply,
directly, to odd lens spaces, in particular to $Z_1$, \ie to the {\it
full} three--sphere. This remark applies to later results too.

We can write the general rotation generating function,
  $$
  g(\si;\Ga)={1+\si^{\de_0}\over(1-\si^{\de_2})(1-\si^{\de_1})}\,,
  \eql{genrot}
  $$
in terms of the degrees $\de_0,\de_1,\de_2$ and can take things
further as in our developments in [\pref{ChandD}]. Simple algebra
gives, from (\peq{gentot}),
  $$
   K^{1/2}(\tau)=-{d\over
   d\tau}{\cosh(\de_0\tau)\over2\sinh(\de_2\tau)\sinh(\de_1\tau)}\,.
   \eql{rootk3}
  $$
I just mention that  a possible direct physical interpretation of
this quantity occurs in thermal field theory on the space--time,
$T\times$S$^3/\Ga'$, because the free energy is given by,
  $$
  F(\be)=E-{1\over\be}\sum_{m=1}^\infty {1\over m} K^{1/2}(m\be)\,,
  \eql{freeen}
  $$
where $\be=1/kT$. Kennedy, [\pref{Kennedy}], gives some discussion of
thermal quantities on this factored Einstein universe.

In (\peq{freeen}), $E$ is the vacuum, zero temperature energy and can
be called the Casimir energy and I now turn to its evaluation. The
numbers were derived in [\pref{DandB}] essentially by direct
substitution of group properties. I now present a more systematic
method.
\section{\bf 4. Casimir energies on spherical factors.}

Since I am concerned, at least initially, with exposing general
techniques, I restrict to a conformally invariant scalar field
theory. In this case, for a freely acting $\Ga'$ on
$T\times$S$^3/\Ga'$, there are no divergences to bother us. As a
consequence, the only other basic result one needs is that the
Casimir energy is given by the value of the \zf\ on S$^3/\Ga'$,
$\ze(s)$, at $s=-1/2$,
  $$
  E={1\over2}\ze\big(-{1\over2}\big)\,.
  \eql{casen}
  $$

The essential calculational point is that the \zf, $\ze(s)$, for the
Laplacian on S$^3/\Ga'$ is related to the \zf\ for the square--root
of the Laplacian by simply,
  $$
  \ze(s)=\ze^{1/2}(2s)\,,
  \eql{zet1}
  $$
and the latter quantity is given by the standard Mellin transform,
  $$
  \ze^{1/2}(s)={1\over\Ga(s)}\int_0^\infty
  d\tau \tau^{s-1}K^{1/2}(\tau),
  \eql{mell1}
  $$
with $K^{1/2}$ as in (\peq{rootk3}). One then has the continuation,
  $$
  \ze(s)={i\Ga(1-2s)\over2\pi}\int_{C}
  d\tau (-\tau)^{2s-1}K^{1/2}(\tau)\,,
  \eql{mell2}
  $$
where $C$ is the Hankel contour.

Looking at (\peq{rootk3}) and integrating by parts gives
  $$
  \ze(s)={i\Ga(2-2s)\over2\pi}\int_{C}
  d\tau (-\tau)^{2s-2}H(\tau)\,,
  \eql{mell3}
  $$
where
  $$
  H(\tau)={\cosh(\de_0\tau)\over2\sinh(\de_2\tau)\sinh(\de_1\tau)}\,.
  \eql{aitch1}
  $$

Note that $H(\tau/2)$ is a square--root heat--kernel on the unit
orbifold S$^2/\Ga$, [\pref{ChandD}], the operator being the conformal
one, ${\bf L^2}+1/4$, with eigenvalues $(2l+1)^2/4$, $l=0,1,\ldots$
and degeneracy $2l+1$.

This means that it is possible to relate the \zfs\ on S$^3/\Ga'$ and
S$^2/\Ga$. Each is given by the general formulae (\peq{zet1}) and
(\peq{mell1}) where, for S$^3/\Ga'$, $K^{1/2}$ is given by
(\peq{rootk3}) while, for the unit S$^2/\Ga$, it equals $H(\tau/2)$,
(\peq{aitch1}) so one has the relation
  $$
  K^{1/2}_{S^3/\Ga'}(\tau)=-{d\over d\tau}K^{1/2}_{S^2/\Ga}(2
  \tau)\,.
  $$
Substitution of this into the previous equations easily yields the
relation,
  $$
  \ze_{S^3/\Ga'}(s)=2^{1-2s}\ze_{S^2/\Ga}(s-1/2)\,,
  \eql{reln}
  $$
the simplest example of which is for $\Ga={\bf 1}$, when the \zfs\
are Riemann, or, better, Hurwitz  ones. The details are mildly
instructive. For $\Ga={\bf 1}$, the doubled group is $\Ga'=Z_2$,
giving the {\it projective} three--sphere. It is well known, (\eg\
Schulman [\pref{Schulman}], [\pref{Dow}]), that $l$ is then
restricted to odd values so
  $$
  \ze_{S^3/Z_2}(s)=\sum_{\rm odd}{n^2\over n^{2s}}\,,
  \eql{3proj}
  $$
and we know that
  $$
  \ze_{S^2}(s)=2^{2s}\sum_{\rm odd}{n\over n^{2s}}\,,
  $$
which confirms (\peq{reln}). Only even lens spaces are accessible via
(\peq{reln}). To avoid misunderstandings, it should be emphasised
that S$^2/\Ga$ refers to an {\it orbifold} quotient. $\Ga$ has fixed
points on the two--sphere.

Equation (\peq{reln}) just reflects the relation, (\peq{diden}),
between the degeneracies, $(2l+1)d_{2l+1}(\Ga')=(2l+1)d(l;\Ga)$, and
can, of course, be deduced immediately from this.

Equation (\peq{reln}) at the point $s=0$  relates the conformal
anomaly in three dimensions to the Casimir energy in two. Both are
zero.

It is now a simple matter to set $s=-1/2$ and evaluate the integral,
(\peq{mell3}), by residues. One finds,
  $$
  E_{\Ga'}= {15\de_0^4  - 30\de_0(\de_1^2  + \de_2^2 ) + 7\de_1^4  + 10\de_1^2\de_2^2
    +7\de_2^4\over720\de_1\de_2}\,,
  \eql{casen3}
  $$
the actual numbers being
  $$
  E_{\bf T'}=-{3761\over8640}\,,\quad E_{\bf O'}=-{11321\over17280}\,,
  \quad E_{\bf Y'}=-{43553\over43200}\,,
  \eql{casen4}
  $$
in agreement with our earlier evaluations, [\pref{DandB}], but
without the rather {\it ad hoc} computations employed there and
outlined in the section 6. It is obvious from the start that the
values are rational. Although $\de_0=\de_1+\de_2-1$, the expressions
are neater if $\de_0$ is retained.

The expression for the dihedral $D_q'$ case is easily obtained from
(\peq{casen3}) as,
  $$
  E_{D_q'}=-{20q^4+8q^2+180q-7\over1440q}\,,
  \eql{casen5}
  $$
and the cyclic $Z_q$ values are,
  $$
  E_{Z_q}=-{q^4+10q^2-14\over720q}\,.
  \eql{casen6}
  $$
\section{\bf 5. Cyclic decompositions.}

Instead of treating each group one by one, labelled by the
corresponding degrees, it is possible, perhaps more economically, to
use the cyclic decompositions (\peq{genfunm}) or (\peq{genfunm2}),
[\pref{Meyer}], p.139, which clearly translate into cyclic
decompositions of the \zfs, [\pref{ChandD}], and thence of the
Casimir energies. For example, from (\peq{genfunm2}) and
(\peq{reln}),
   $$
  \ze_{\Ga'}(s)
  ={1\over2}\bigg(\sum_q \ze_{Z_{2q}}(s)-\ze_{Z_2}(s)\bigg)\,,
  \eql{zedecomp}
  $$
and so, in particular,
   $$
  E_{\Ga'}
  ={1\over2}\bigg(\sum_q E_{Z_{2q}}-E_{Z_2}\bigg)\,,
  \eql{edecomp}
  $$
which works for ${\bf T',\,O'}$ and ${\bf Y'}$ using (\peq{casen6}).

The \zf, (\peq{mell3}), is related to the Barnes \zf\ employed in
earlier works. In view of (\peq{reln}) we can equivalently repeat the
formula for the two--sphere case, [\pref{ChandD}],
  $$
  \ze_{S^2/\Ga}(s)=\ze_2(2s,1/2\mid \de_1,\de_2)
  +\ze_2(2s,\de_1+\de_2-1/2\mid \de_1,\de_2)\,,
  \eql{barnes1}
  $$
where, generally,
  $$\eqalign{
  \ze_d(s,a\mid \bom)&={i\Ga(1-s)\over2\pi}\int_C d\tau
  {\exp(-a\tau)(-\tau)^{s-1}\over\prod_{i=1}^d\big(1-\exp(-\om_i\tau)
\big)}\cr &=\sum_{{\bf m}={\bf 0}}^\infty{1\over(a+\bom{\bf.m})^s}\,.
 }
 \eql{barnes2}
  $$

The residues and values of the Barnes function are given in terms of
generalised Bernoulli functions,  of which (\peq{casen3}) is an
example and it is clear that this whole process can easily be
automated and extended  to the higher spheres.
\section{6. \bf Cosecant sums.}

From the basic definition, (\peq{czeta}), the one--sided 3--sphere
\zf\ emerges directly as a sum of derivatives of Epstein \zfs,
[\pref{DandB}],
  $$
  \ze(s)=-{1\over2|\Ga'|}\sum_\ga{1\over\sin\th_\ga}{\pa\over\pa\th_\ga}
  Z\bigg|\matrix{0\cr\th_\ga/2\pi}\bigg|(2s)\,,
  \eql{ep1}
  $$
where $Z$ is the simplest Epstein function (it has other names),
   $$
   Z\bigg|\matrix{0\cr\th/2\pi}\bigg|(2s)=\sumdasht{-\infty}{\infty}
   {e^{in\th}\over n^{2s}}\,.
   \eql{ep2}
   $$
I will denote it by $Z_E(\th,s)$, for short.

From this expression an alternative form of the Casimir energy was
derived in [\pref{DandB}]. From the standard formula
  $$
  2\sum_{l=1}^\infty\sin l\th=\cot(\th/2)\,,
  \eql{form1}
  $$
one finds
  $$
  E={1\over|\Ga'|}\bigg[{1\over240}-{1\over16}\sum_{\ga\ne{\bf 1}}
  \cosec^4\big({\th_\ga\over2}\big)\bigg]\,.
  \eql{svacen1}
  $$

More generally, in the two--sided case,
  $$
  E={1\over|\Ga'|}\bigg[{1\over240}-{1\over16}\sum_{\ga\ne{\bf 1}}
  \cosec^2\big({\al\over2}\big)\cosec^2\big({\be\over2}\big)\bigg]\,,
  \eql{svacen2}
  $$
in terms of the angles (\peq{angles2}). For right actions only,
$\al=\be=\th_R=\th_\ga$.

For lens and prism spaces one can use standard, and often very old
(some dating to Euler) finite sums of powers of cosecants to give
polynomials in $q$, in agreement with the results stated earlier. For
the other groups, direct substitution of the angles yielded the
values in (\peq{casen4}) after cancellations. See also the
computations in Gibbons {\it et al} [\pref{GPR}].

Of course, some derivations of these cosecant sums boil down to
residue evaluations and so these sums in themselves are somewhat of a
detour. A brief history was attempted in [\pref{Dow7}] and more
references can be found in Berndt and Yeap, [\pref{BandY}].
\mgn{CHECK} Some explicit expressions are given later in connection
with the corresponding spinor calculation.

From the purely numerical aspect, an Epstein approach is made more
attractive by the existence of an exponentially convergent series
involving the incomplete $\Ga$--function, $\Ga(s,a)$, for which there
is a rapid, continued fraction algorithm. Against this must be set
the fact that the angles $\th_\ga$ have to be individually put in.

We have used this method before. Here, I consider its use for the
evaluation of $\ze'(0)$. The relevant expression is, [\pref{DandJ}],
  $$\eqalign{
  \pi^{-s}\Ga(s)&{\pa\over\pa\th}Z_E(\th,s)\cr
  &=-2\sum_{n=1}^\infty n\sin (n\th)\,\Ga(s,\pi n^2)
  -\sum_{n=-\infty}^\infty(n+h)\Ga\big((3-2s)/2,\pi(n+h)^2\big)
  }
  \eql{igam1}
  $$
with $h=\th/2\pi$. The transformations leading to this expression are
already in Epstein, [\pref{Epstein}].

An important analytical fact about this formula is that it has
exactly the combination needed to compute $\ze'(0)$. To see this we
need only note that
  $$
  \lim_{s\to0}\Ga(s)f(s)\sim f'(0)+f(0)\big({1\over s}+\ga\big)
  $$
and that $Z_E(\th,0)=0$. (There is no conformal anomaly on
S$^3/\Ga'$.) Therefore one has quite simply,
  $$\eqalign{
  &{\pa\over\pa\th}Z'_E(\th,0)\cr
  &=-2\sum_{n=1}^\infty n\sin (n\th)\,\Ga(0,\pi n^2)
  -\sum_{n=-\infty}^\infty(n+h)\Ga\big(3/2,\pi(n+h)^2\big)\,,
  }
  \eql{igam2}
  $$
which can be substituted into
  $$\eqalign{
  \ze'(0)&=-{1\over2|\Ga'|}\sum_\ga{1\over\sin\th_\ga}{\pa\over\pa\th_\ga}
  Z'_E(\th_\ga,0)\cr
  &={2\over|\Ga'|}\bigg(\ze_R'(-2)-{1\over4}
  \sum_{\ga\ne{\bf1}}{1\over\sin\th_\ga}{\pa\over\pa\th_\ga}
  Z'_E(\th_\ga,0)\bigg)\,,
  }
  \eql{ep3}
  $$
and the sum over $\ga\ne{\bf1}$ for ${\bf T}$, ${\bf O'}$ and ${\bf
Y'}$ performed angle by angle, as mentioned before.

This expression is not pursued here because another route to this
quantity is given in the next section.

\section{7. \bf Functional determinants.}

Formula (\peq{ep3}) allows one to compute the Laplacian determinant,
$\exp\big(-\ze'(0)\big)$. Alternatively, equations (\peq{reln}) and
(\peq{barnes1}) mean that it is possible to find expressions for the
functional determinants on the factored three--sphere in terms of the
Barnes function, and thence, if desired, of the Hurwitz \zf, which is
often how these answers are left. In [\pref{Dow3, Dow9}] we have
discussed such questions and again can make use of this work here. Of
course, there are many other relevant references, but this is not a
historical work.

From the relation (\peq{reln}) one gets,
  $$
  \ze'_{S^3/\Ga'}(0)=2\ze'_{S^2/\Ga}(-1/2)\,,
  \eql{reln2}
  $$
where I have used the vanishing of $\ze_{S^3/\Ga'}(0)$, at least for
conformal scalars and spinors. From (\peq{barnes1}) it is seen that
one is required to evaluate
  $$
  \ze'_2(-1,a\mid \de_1,\de_2)\,,\quad (a=1/2,\,\,\de_1+\de_2-1/2)
  $$
and the problem devolves upon computing the derivative of the Barnes
function at negative integers. This has been treated in
[\pref{Dow3,Dow9}]. The analysis in [\pref{Dow3}] allows one to
obtain `exact' expressions in terms of derivatives of the `lower'
Hurwitz \zf. The procedure involves breaking up the summation over
${\bf m}$ in the Barnes function, (\peq{barnes2}), using residue
classes.

 Rather than treat general degrees, it is somewhat easier to calculate
$\ze'(0)$ on lens spaces and then use the cyclic decomposition,
(\peq{zedecomp}). The relation (\peq{reln2}) specialises to
   $$
  \ze'_{S^3/Z_{2q}}(0)=2\ze'_{S^2/Z_q}(-1/2)\,.
  \eql{reln3}
  $$

The degrees for the lens case
are $\de_1=q$, $\de_2=1$. Using residue classes mod $q$, manipulation of
the sum definition of the Barnes function yields the expression
  $$
  \ze_{S^2/Z_q}(s)={2\over q}\,\ze_R\big(2s-1,{1\over2}\big)+
  \ze_R\big(2s,{1\over2}\big)-
  {1\over q^{2s+1}}\sum_{p=0}^{q-1}(2p+1)\ze_R\big(2s,{2p+1\over2q}\big)
  \eql{rotzet}
  $$
which was referred to as the orbifolded S$^2$ {\it rotational} \zf\
in [\pref{ChandD}] and was used in [\pref{Dow3}] to compute
two--sphere determinants. Here one requires the value of the
derivative at $s=-1/2$,
  $$\eqalign{
   \ze'_{S^2/Z_q}\big(-{1\over2}\big)&=
   2\ze'_R\big(-1,{1\over2}\big)
  +{4\over q}\ze'_R\big(-2,{1\over2}\big)+
  2\log q\sum_{p=0}^{q-1}(2p+1)\ze_R\big(-1,{2p+1\over2q}\big)\cr
  &\hspace{*******************}-
  2\sum_{p=0}^{q-1}(2p+1)\ze'_R\big(-1,{2p+1\over2q}\big)\cr
  &={1\over12}\log(q/2)-\ze_R'(-1)-{3\over q}\ze'_R(-2)-
  2\sum_{p=0}^{q-1}(2p+1)\,\ze'_R\big(-1,{2p+1\over2q}\big)\cr
  }
  \eql{deriv1}
  $$
which could, possibly, be thought of as `exact' but is, at least, in
a form suitable for numerical treatment, \cf\ Nash and O'Connor,
[\pref{NandO}]. In the derivation of this formula further use has
been made of the fact that $\ze_{S^2/Z_q}(-1/2)$ is zero.

It is possible to find an alternative expression for the \zf\ that
displays this vanishing and allows the derivatives to be avoided, in
analogy to the result $\ze'_R(-1)=-\ze_R(3)/4\pi^2$. The details are
given in [\pref{ChandD}] that give rise to the alternative form,
  $$\eqalign{
  \ze_{S^2/Z_q}(s)={2\over q}\,\ze_R&\big(2s-1,{1\over2}\big)+
  {2^{2s}\Ga(1-2s)\cos\pi s\over q\pi^{1-2s}}\cr
  &\times\sum_{p=1}^{q-1}{1\over\sin(\pi p/q)}
  \bigg(\ze_R(1-2s,{p\over q})-2^{2s}\ze_R(1-2s,{p+q\over
  2q})\bigg)\,,
  }
  \eql{altrotzet}
  $$
showing the zeros at $s=-(2k+1)\pi/2$, with $k=0,1,\ldots$. Despite
appearances, the only pole is at $s=1$, correctly.

The required derivative follows as the numerically easier formula,
  $$
  \ze'_{S^2/Z_q}\big(-{1\over2}\big)={3\over4\pi^2q}\,\ze_R(3)
  +{1\over2q\pi }\sum_{p=1}^{q-1}{1\over\sin(\pi p/q)}
  \bigg(\ze_R(2,{p\over q})-{1\over2}\ze_R(2,{p+q\over 2q})\bigg)\,.
  $$

In order to cover ${\bf T'}$, ${\bf O'}$ and ${\bf Y'}$ the values
$q=1,2,3,4$ and $5$ are needed, the simple $q=1$ case being already
given in (\peq{3proj}).

I present the numbers for the scalar (conformal) determinant,
$\det=e^{-\ze'(0)}$, on S$^3/\Ga'$,
 $$\eqalign{
  \det({\bf T'})&=0.2020887\cr
  \det({\bf O'})&=0.1287757\cr
  \det({\bf Y'})&=0.0730560\,,\cr
}
 $$
and display a graph of $W=-\log\det$ for the even, one--sided lens
spaces. The determinant tends to zero as $q\to\infty$.
\vskip0.3truein
\input epsf
\epsfbox{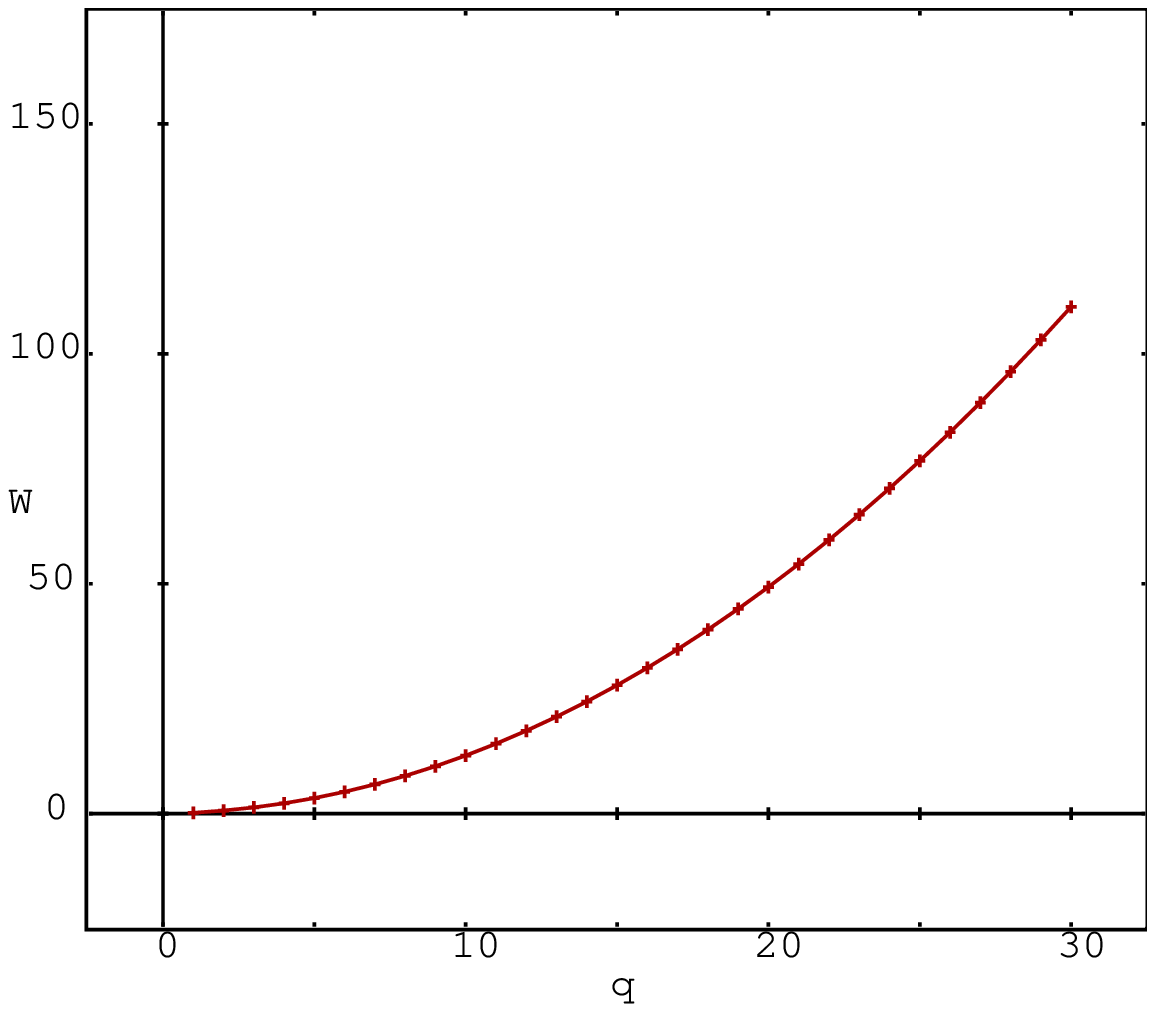}

{fig1. $W=-\log\det$ for conformal scalars on lens spaces of order
2q}
\begin{ignore}
In order that the resulting sums should be as small as possible, it
is advantageous to remove any common factors between $\om_1$ and
$\om_2$ and write $\om_1=ce_1$, $\om_2=ce_2$ where $e_1$ and $e_2$
are coprime.

Then set
  $$
  m_1=n_1e_2+p_2\,,\quad m_2=n_2 e_1+p_1\,,
  $$
with $0\le p_i<e_i-1$ and $n_i=0,1,\ldots,\infty$. The denominator in
(\peq{barnes2}) then equals $c(b+e_1e_2(n_1+n_2)+p_2e_1+p_1e_2)$
where $b=a/c$, and the summation can be re--arranged
  $$\eqalign{
  \ze_2(s,a\mid \bom)&=c^{-s}\sum_{\bf p}\sum_{n=0}^\infty{1+n\over
  (b+e_1e_2n+p_2e_1+p_1e_2)^s}\cr
  &=\bigg({c\over\om_1\om_2}\bigg)^s\bigg[\sum_{{\bf p},n}
  {1\over (n+w_b)^{s-1}}+
  \sum_{\bf p}(1-w_b)\sum_{n=0}^\infty{1\over (n+w_b)^s}\bigg]\cr
  &=\bigg({c\over\om_1\om_2}\bigg)^s\sum_{\bf p}\big(\ze_R(s-1,w_b)
  +(1-w_b)\ze_R(s,w_b)\big)\,,
  }
  \eql{barnes3}
  $$
with
  $$
  w_b={b\over e_1e_2}+{p_1\over e_1}+{p_2\over e_2}\,.
  $$
There is a more efficient restructuring but this would not be of
great advantage here.

(\peq{barnes3}) presents the two--dimensional integral Barnes
function as a finite series of Hurwitz \zfs\ and we can see from
(\peq{zedash0}) what the structure of $\ze'_{S^3/\Ga'}(0)$ will be as
a result. For the derivative one gets
  $$\eqalign{
  \ze'_2(-n,a\mid \bom)&=
  \bigg({\om_1\om_2\over c}\bigg)^n\sum_{\bf p}\big(\ze'_R(-n-1,w_b)
  +(1-w_b)\ze'_R(-n,w_b)\big)\cr
  &+\bigg({\om_1\om_2\over c}\bigg)^n\log({c\over\om_1\om_2})
  \sum_{\bf p}\big(\ze_R(-n-1,w_b)
  +(1-w_b)\ze_R(-n,w_b)\big)\,.
  }
  \eql{zedash2}
  $$
The second sum can be made quite explicit noting the standard values
of the Hurwitz function
  $$
  \ze_R(-n,w)=-{B_{n+1}(w)\over n+1}\,,
  $$
in terms of the simple Bernoulli polynomials. We are interested in
$n=1,2$ so that the second summand is
  $$
  -{1\over3}B_3(w)+(1-w){1\over2}B_2(w)=
  -{1\over12}\big(10w^3-18w^2+9w-1\big)\,,
  \eql{secsum}
  $$
and the summations over ${\bf p}$ are easily accomplished. The
general expression is not especially helpful. It is only necessary to
do the arithmetic, noting that, since $\om_1=2\de_1$ and
$\om_2=2\de_2$, $\om_1\om_2=4|\Ga|$.

As a specific example I set up the octahedral case for which
$\de_1=4,\de_2=6$ so $e_1=2, e_2=3$ and $c=4$. In the first term of
(\peq{zedash0}), $b=1/4$, and, in the second, $b=19/4$. Also
  $$
  w_b={b\over6}+{p_1\over2}+{p_2\over3}\,,\quad {\rm and}\quad
  \sum_{\bf p}=\sum_{p_1=0}^1\sum_{p_2=0}^2
  $$

The second term in (\peq{zedash2}), from (\peq{secsum}), gives the
contribution to (\peq{zedash0})
  $$
  -{7715\over32}\log(1/24)\approx766.21
  $$
\end{ignore}
\section{\bf 8. Spinors.}

To analyse the astrophysical data one needs only the scalar
harmonics. However it is within our scope to consider other fields
and I now lay out some comments on the Dirac field without going into
too many details as most of these are available elsewhere.

The eigenproblem for spin--half on spheres, and therefore on the
Einstein universe, is well known, going back at least to
Schr\"odinger. Basic facts are that the eigenvalues of the squared,
massless Dirac operator are given by
  $$
  \la_n={1\over a^2}(n+1/2)^2\,,\quad n=1,2,\ldots
  \eql{deigen}
  $$
with degeneracies $2n(n+1)$, for a two--component field.

On S$^3/\Ga'$ the total degeneracies (left times right) are,
  $$
  D_n(\Ga')={1\over|\Ga'|}\sum_\ga\big((n+1)\chi_n(\th_\ga)
  +n\chi_{n+1}(\th_\ga)\big)\,.
  \eql{ddegen}
  $$
The two parts to $D_n$ correspond to the fact that the positive and
negative eigenvalues of the Dirac operator have been combined into
(\peq{deigen}). Look at the two parts in turn using the previous
analysis of the right degeneracy (\peq{degen3}) on S$^3/\Ga'$. The
first part is zero unless $n$ is odd and the second is zero unless
$n$ is even, see (\peq{degen6}). (I am again excluding odd lens
spaces.) Thus in the first part, set $n=2l+1$, and in the second
$n=2l+2$ with $l=0,1,\ldots$ in both cases. Hence, from
(\peq{diden}),
  $$
   D_{2l+1}(\Ga')=(2l+2)\,d(l;\Ga)\,,
   \quad D_{2l+2}(\Ga')=(2l+2)\,d(l+1;\Ga)\,,
   \eql{ddegen2}
  $$
in terms of the S$^2/\Ga$ degeneracies. Our previous formulae, \eg\
(\peq{restdegen}), can be used to make (\peq{ddegen2}) more explicit.

The heat--kernel for the (positive) square--root of the squared
massless Dirac operator on S$^3/\Ga'$ is then, \cf (\peq{rootk}),
  $$\eqalign{
   K_S^{1/2}(\tau)&=\sum_{n=1}^\infty D_n(\Ga')\,e^{-(n+1/2)\tau}\cr
   &=\sum_{l=0}^\infty(2l+2)d(l;\Ga)e^{-(2l+3/2)\tau}+
   \sum_{l=0}^\infty(2l+2)d(l+1;\Ga)e^{-(2l+5/2)\tau}\cr
   &=-e^{\tau/2}{d\over d\tau}\sum_{l=0}^\infty
   d(l;\Ga)e^{-(2l+2)\tau}-e^{-\tau/2}{d\over d\tau}\sum_{l=0}^\infty
   d(l+1;\Ga)e^{-(2l+2)\tau}\cr
   &=-e^{\tau/2}{d\over d\tau}e^{-2\tau}\sum_{l=0}^\infty
   d(l;\Ga)e^{-2l\tau}-e^{-\tau/2}{d\over d\tau}\sum_{l=0}^\infty
   d(l+1;\Ga)e^{-(2l+2)\tau}\cr
   &=-e^{\tau/2}{d\over d\tau}e^{-2\tau}\sum_{l=0}^\infty
   d(l;\Ga)e^{-2l\tau}-e^{-\tau/2}{d\over d\tau}\sum_{l=0}^\infty
   d(l;\Ga)e^{-2l\tau}\,,\cr
   }
   \eql{spk}
  $$
where a zero term has been added to the second sum. One can set
$\si=e^{-2\tau}$ in order to make contact with the SO(3) generating
functions (\peq{genfunm}) which one notes from [\pref{ChandD}] are
related to the S$^2/\Ga$ Laplacian square root heat--kernels,
$H(\tau/2)$, (\peq{aitch1}), by,
  $$
  g(\si;\Ga)=e^{\tau} H(\tau)
  \eql{gh}
  $$
and so
  $$\eqalign{
  K_S^{1/2}(\tau)&=-e^{\tau/2}{d\over d\tau}\,e^{-\tau}H
  -e^{-\tau/2}{d\over d\tau}\,e^\tau H\cr
  &=-2\sinh(\tau/2)\,H-2\cosh(\tau/2)\,{d\over d\tau}H\,.
  }
  \eql{sphk}
  $$

The spinor \zf\ is given by the general formula (\peq{mell2}) with
(\peq{sphk}) and the derivative can again be removed by an
integration by parts yielding two integrals,
  $$\eqalign{
  \ze_S(s)=&{i\Ga(2-2s)\over\pi}\int_Cd\tau(-\tau)^{2s-2}
  \cosh(\tau/2)\,H(\tau)\cr
  &-{i\Ga(1-2s)\over2\pi}\int_Cd\tau(-\tau)^{2s-1}
  \sinh(\tau/2)\,H(\tau)\,.\cr
  }
  \eql{spzet2}
  $$
We can confirm from this that $\ze_S(0)=0$.

For the two--component spinor Casimir energy,
  $$
   E_{\Ga'}=-{1\over2}\ze_S\big(\!-\!{1\over2}\!\big)\,,
   \eql{vacensp}
  $$
a standard residue evaluation gives,
  $$\eqalign{
  E_{\Ga'}=&{1\over 5760\de_1\de_2}\bigg(128\de^4_1 + 128\de^4_2
  - 640\de_1^2 \de^2_2+ 1920\de^2_1 \de_2 +1920\de_1\de^2_2 \cr
  &  - 4320\de_1\de_2  - 1440\de^2_1 - 1440\de^2_2  + 2400\de_1+
  2400\de_2 - 1005\bigg)\,.
  }
  \eql{spvac}
  $$
In particular,
   $$\eqalign{
  E_{Z_{2q}}&={128q^4  - 160q^2  + 83\over 5760q}\cr
  E_{D'_q}&={128q^4  - 160q^2  + 1440q + 83\over11520q}\cr
   E_{\bf T'}&={40211\over69120}\,,\quad
   E_{\bf O'}={135251\over138240}\,,\quad
   E_{\bf Y'}={567443\over345600}\,.\cr
  }
  \eql{spvacv}
     $$
The last values can also be obtained from the cyclic decomposition,
(\peq{edecomp}), which is true generally.

\section{\bf 9. The angle form.}

As a check, I derive the spinor equivalent of (\peq{svacen1}), which
can be called the image form of the vacuum energy. This follows on a
direct evaluation of the original summation expression for the \zf,
  $$
  \ze(s)=\sum_n{d_n\over\la_n^s}=
{1\over|\Ga'|}\sum_\ga\sum_{n=1}^\infty{1\over(n+1/2)^{2s}}\,
\big((n+1)\chi_n(\th_\ga)
  +n\chi_{n+1}(\th_\ga)\big)\,.
  \eql{spzet}
  $$
The only divergent term is for the identity, $\ga=E={\rm id}$ ,
$\th_\ga=0$, which is easily treated by continuing to a Hurwitz \zf.
  $$\eqalign{
  \ze_E(s)&={2\over|\Ga'|}\sum_{n=1}^\infty{n(n+1)\over(n+1/2)^{2s}}\cr
  &={2\over|\Ga'|}\big(\ze_R(2s-2,1/2)
  -{1\over4}\ze_R(2s,1/2)\big)\,,
  }
  \eql{idterm}
  $$
a very old expression as, apart from the volume factor, $|\Ga'|$ this
is just the full sphere result. It could be rearranged in several
inessential ways. For example, at the sum level, one can introduce
$\overline n= 2n+1$ and rewrite the sum over odds as all minus evens.

The other terms, $\ga\ne E$, do not diverge at $s=-1/2$, and I can
proceed directly with the sum as it stands. The Casimir energy,
(\peq{vacensp}), is
  $$\eqalign{
  E_{\Ga'}&=-{1\over|\Ga'|}
  \bigg(\ze_R(-3,1/2)
  -{1\over4}\ze_R(-1,1/2)\bigg)-{1\over|\Ga'|}\sum_{\ga\ne
  E}\sum_{n=1}^\infty\big(n^2+{1\over2}\big)\chi_n(\th_\ga)\cr
  &={1\over8|\Ga'|}\bigg(
  {17\over120}-\sum_{\ga\ne E}\big(
  \cosec^2\th_\ga/2-\cosec^4\th_\ga/2\big)\bigg)\,,
  }
  \eql{spvacang}
  $$
using (\peq{form1}).

 The $\cosec^4$ sum is that occurring in the scalar vacuum energy
 and the $\cosec^2$ part is a novelty occasioned by the spectral
 asymmetry of the Dirac operator on a factored space.

 Defining the cosecant sums,
   $$
   C(r;\Ga')={1\over|\Ga'|}\sum_{\ga\ne E}\cosec^{2r}\th_\ga/2\,,
   \eql{cosecs}
  $$
brute force angle substitution gives
  $$\eqalign{
  C(1;{\bf T'})&={167\over72}\,,\quad C(2;{\bf T'})
  ={1505\over216}\cr
   C(1;{\bf O'})&={383\over144}\,,\quad C(2;{\bf O'})
   ={4529\over432}\cr
   C(1;{\bf Y'})&={1079\over360}\,,\quad \!\!C(2;{\bf Y'})
   ={87109\over5400}\,,\cr
  }
  \eql{cosecvs}
  $$
and the old summations mentioned earlier are
  $$\eqalign{
  C(1;Z_q)&={q^2-1\over3q}\,,\quad C(2;Z_q)={(q^2+11)(q^2-1)\over 45q}\cr
  C(1;D'_q)&={4q^2+12q-1\over12q}\,,\quad C(2;D'_q)=
  {16q^4+40q^2+360q-11\over180q}\,,\cr
   }
   \eql{cosecvs2}
  $$
the last sum by Jadhav, in this way. Combining these values yields
the spinor Casimir energies, (\peq{spvacv}), previously obtained by
the  alternative method involving the degrees.\mgn{DO CYCLIC,
DIHEDRAL}
\section{\bf 10. The Maxwell field.}

To complete the set of standard fields I now consider massless
spin--one. Actually, it is possible to treat all three spins, $0$,
$1/2$ and $1$, together [\pref{Dow10}], but, for transparency, it has
been decided to keep them apart.

The solution of Maxwell equations on the Einstein Universe is well
known and again goes back to Schr\"odinger. I deal with transverse
fields. The eigenvalues of the square of the first order $\curl$
operator are
  $$
  \la_n=n^2\,,\quad n=2,3,\ldots
  \eql{meig}
  $$
with degeneracies $d_n=2(n^2-1)$ on the full sphere. On the factored
sphere, the (total) degeneracies are,
  $$
  d_n(\Ga')={1\over|\Ga'|}\sum_\ga
  \big((n+1)\chi_{n-1}(\th_\ga)+(n-1)\chi_{n+1}(\th_\ga)\big)
  \eql{mdegen}
  $$
and again the existence of two parts can be ascribed to a spectral
asymmetry.

For Maxwell theory, there is a gauge question. In addition to the
transverse field, (\ie coexact 1--form) one must subtract a harmonic
zero form. One way of doing this, formally, on the full sphere is to
extend and double up the summation range. For the \zf,
  $$
  \ze(s)=\sumdasht{-\infty}{\infty} {n^2-1\over n^{2s}}
  =2(\ze_R(2s-2)-\ze_R(2s))\,.
  \eql{vzetf}
  $$

Although it seems nothing has been done, the value $\ze(0)=1$ can now
be interpreted as a consequence of the ghost zero mode and {\it not}
as an indication of a constant term in the expansion of the \hk,
[\pref{Dow8}]. These considerations can be dispensed with if one is
concerned just with the vacuum, zero--point energy. They would come
into play for functional determinants but these are left for another
time.

Returning to (\peq{mdegen}), the previous analysis shows that $n$
must be even, $n=2l+2$, and so for the Maxwell cylinder heat--kernel
  $$\eqalign{
  K^{1/2}_M(\tau)
  &=\sum_{l=0}^\infty\big((2l+3)d_l(\Ga)+(2l+1)d_{l+1}(\Ga)\big)
  e^{-(2l+2)\tau}\cr
  &=e^\tau{d\over d\tau}\,e^{-2\tau}H(\tau)+
  e^{-\tau}{d\over d\tau}\,e^{2\tau}(H(\tau)-e^{-\tau})
  }
  \eql{mcylin}
  $$
using (\peq{genfunm}), (\peq{gh}). The corresponding \zf\ is,
 $$\eqalign{
  \ze_M(s)=&{i\Ga(2-2s)\over2\pi}\int_Cd\tau(-\tau)^{2s-2}
  (2\cosh\tau\,H(\tau)-1)\cr
  &-{i\Ga(1-2s)\over2\pi}\int_Cd\tau(-\tau)^{2s-1}
  (2\sinh\tau\,H(\tau)-1)\,.\cr
  }
  \eql{mzet2}
  $$
A simple check is the value $\ze_M(0)=1$ which arises from the ``1"
term in the second integrand and which has a zero mode connotation.
One confirms that it does not contribute to the residue when
evaluating $\ze_M(-1/2)$ and finds for the Casimir energy,
  $$\eqalign{
  E_{\Ga'}=
   -{1 \over90\de_1\de_2}\,(2\de_1^4 + 2\de^4_2& - 10\de_1^2 \de^2_2
     + 30\de_1^2 \de_2   + 30\de_1\de^2_2\cr
        &- 135\de_1\de_2- 45\de \de^2_1 - 45\de^2_2 + 105\de_1 + 105\de_2 -
        60)\,.
    }
    \eql{mvac}
   $$
Explicit values are,
   $$\eqalign{
   E_{Z_{2q}}&=-{2q^4  - 25q^2  + 2\over90q}\cr
   E_{D'_q}&=-{2·q^4  - 25·q^2  - 45·q + 2\over180q}\cr
   E_{\bf T'}&={79\over270}\,,\quad
   E_{\bf O'}={23\over1080}\,,\quad
   E_{\bf Y'}=-{698\over1350}\,.\cr
  }
  \eql{mvacv}
     $$
The Maxwell vacuum energy is negative on dodecahedron space.

The angle form can again be produced as a check, and for interest. We
have, [\pref{DandJ}],
  $$\eqalign{
  E_{\Ga'}&={1\over|\Ga'|}
  \bigg(\ze_R(-3,1)-\ze_R(-1,1)\bigg)+{1\over|\Ga'|}\sum_{\ga\ne
  E}\sum_{n=1}^\infty\big(n^2+2\big)\chi(\th_\ga)\cr
  &={1\over2|\Ga'|}\bigg(
  {11\over60}+\sum_{\ga\ne E}\big(
  \cosec^2\th_\ga/2-{1\over4}\,\cosec^4\th_\ga/2\big)\bigg)\,,
  }
  \eql{mvacang}
  $$
and, of course, calculation gives agreement with (\peq{mvacv}).
\section{\bf 11. Spectral asymmetry.}

As another example of the use of the eigenvalue expressions I will
derive expressions for the spectral asymmetry quantity that occurs as
a boundary correction to the index theorem, applied to four
dimensions. Textbook discussions concern the Dirac equation, the
signature and the de Rham complex, \eg\ [\pref{Gilkey}], and the
corresponding literature is extensive. My treatment of context and
content will be brief.

The Atiyah--Patodi--Singer spectral asymmetry function $\eta(s)$ is
  $$
  \eta(s)=\sumdash{\la}{({\rm sign}\,\la)\over|\la|^s}\,.
  $$

Restricting to one--sided quotients, the construction of $\eta$
corresponds, effectively, to changing the sign of the second,
negative spectrum part of (\peq{ddegen}), (\peq{mdegen}) or of
(\peq{spk}), (\peq{mcylin}) and also setting $2s\to s$. Following
this through gives
   $$\eqalign{
  \eta_S(s)=&-{i\Ga(2-s)\over\pi}\int_Cd\tau(-\tau)^{s-2}
  \sinh(\tau/2)\,H(\tau)\cr
  &+{i\Ga(1-s)\over2\pi}\int_Cd\tau(-\tau)^{s-1}
  \cosh(\tau/2)\,H(\tau)\,,\cr
  }
  \eql{speta}
  $$
for spin--half and
  $$\eqalign{
  \eta_M(s)=&-{i\Ga(2-s)\over2\pi}\int_Cd\tau(-\tau)^{s-2}
  (2\sinh\tau\,H(\tau)-1)\cr
  &+{i\Ga(1-s)\over2\pi}\int_Cd\tau(-\tau)^{s-1}
  (2\cosh\tau\,H(\tau)-1)\,,\cr
  }
  \eql{meta}
  $$
in the Maxwell case. It can again be seen that
$\eta\big(-(2n+1)\big)=0$ from the vanishing of any residues.
Furthermore other values can be readily found. Consider $\eta(-2n)$
and start with $\eta(0)$. Straightforward computation of residues
yields
  $$\eqalign{
  \eta_S(0)&=
 {4\de^2_1 + 4\de^2_2 + 12\de_1\de_2   - 12\de_2 -12\de_1
 + 7\over 12\de_1\de_2}\cr
 \eta_M(0)&={2\de^2_1 + 2\de^2_2  + 3\de_1\de_2  - 6\de_2-6\de_1
 + 5\over 3\de_1\de_2}
  }
  \eql{eta0s}
  $$
and one can, once more, avoid the angle substitution  employed, in
this context, by Gibbons {\it et al}, [\pref{GPR}]. The results agree
in detail with the values in this reference some of which we repeat,
  $$
  \eta_{\bf T'}={167\over144}\,,\quad\eta_{\bf O'}={383\over288}\,,
  \quad\eta_{\bf Y'}={1079\over720}\,.
  $$

The cyclic decomposition could also have been employed, and I will
now do so for the other values of $\eta$ by specialising to even lens
spaces, $L(2q;1,1)$. By putting in the particular expression for $H$,
(\peq{aitch1}), one has
  $$\eqalign{
  \eta_S(s)=&-{i\Ga(2-s)\over4\pi}\int_Cd\tau(-\tau)^{s-2}
  {\coth q\tau\over\cosh(\tau/2)}\cr
  &+{i\Ga(1-s)\over8\pi}\int_Cd\tau(-\tau)^{s-1}
  {\coth q\tau\over\sinh(\tau/2)}\,,\cr
  }
  \eql{speta2}
  $$
for spin--half and
  $$\eqalign{
  \eta_M(s)=&-{i\Ga(2-s)\over2\pi}\int_Cd\tau(-\tau)^{s-2}
  (\coth q\tau-1)\cr
  &+{i\Ga(1-s)\over2\pi}\int_Cd\tau(-\tau)^{s-1}
  (\coth q\tau\coth \tau-1)\,,\cr
  }
  \eql{meta2}
  $$
in the Maxwell case. Standard expansions allow one to write, $n>0$,
  $$\eqalign{
  \eta_S(-2n)&={2^{-2n-4}\over n+1}\sum_{m=0}^{n+1}\comb{2n+2}{2m}2^{4m}
  B_{2m}\bigg(E_{2n-2m+2}+{D_{2n-2m+2}\over2n+1}\bigg)q^{2m-1}\cr
  \eta_M(-2n)&={2^{2n+2}\over2n+1}\bigg(B_{2n+2}\,q^{2n+1}+\!\!
  {1\over
  2n+2}\sum_{m=0}^n\comb{2n+2}{2m}B_{2m}B_{2n-2m+2}q^{2m-1}\bigg)\,.
  }
  \eql{etan}
  $$
$E_n$ are Euler numbers and the $D_n$  are related to the Bernoulli
numbers by $D_n=2(1-2^{n-1})B_n$. The expressions are related to the
expansion coefficients of the relevant heat--kernel, [\pref{ChandD}].

These and earlier results are derived on the assumption that $q$ is
even. They can be extended to odd lens spaces by setting
$2q=\overline q$ when they will apply to S$^3/Z_{\overline q}$ for
all $\overline q$.

Some particular values are
   $$\eqalign{
  \eta_S(0)&={1\over6\ol q}(\ol q^2-1)\cr
   \eta_S(-2)&={1\over360\ol q}(\ol q^2 - 1)(4\ol q^2  + 29)\cr
  \eta_S(-4)&={1\over10080\ol q}(\ol q^2 - 1)(48\ol q^4  + 272\ol q^2  + 1609)\cr
  }
  $$
and
  $$\eqalign{
  \eta_M(0)&={1\over3\ol q}(\ol q-1)(\ol q-2)\cr
  \eta_M(-2)&={1\over45\ol q}(\ol q^2-1)(\ol q^2-4)\cr
  \eta_M(-4)&={1\over45\ol q}(\ol q^2-1)(\ol q^2-4)(3\ol q^2+8)\cr
  \eta_M(-6)&={1\over315\ol q}(\ol q^2-1)(\ol q^2-4)
  (3\ol q^4+10\ol q^2+24)\cr
   \eta_M(-8)&={1\over1260\ol q}(\ol q^2-1)(\ol q^2-4)
  (25\ol q^6+92\ol q^4+272\ol q^2+640)\,.\cr
  }
  $$
These results exhibit the fact that $\eta_M(s)$ vanishes on the full
sphere ($\ol q=1$) and on the projective sphere, $L(2;1,1)$. The
latter fact follows immediately from the angle sum form of Atiyah,
Patodi and Singer, since the only $\th_\ga=\pi$. It also can be seen
in the contour integral forms, (\peq{meta2}), (\peq{speta2}). Note
that only this lens space retains the full, global symmetry of S$^3$.
The spin--one $\eta$ is really the {\it signature}, which vanishes
when there is an orientation preserving isometry, as on the
projective sphere, see \eg\ Hanson and R\"omer, [\pref{HandR}].

The Maxwell $\eta_M(-2n)$, (\peq{etan}), was derived by ourselves
some time ago using the more involved techniques in (\pref{DandJ})
and (\pref{Jadhav}). It can be rearranged using an identity of
Apostol, [\pref{Apostol}], in terms of a generalised Dedekind sum,
[\pref{ChandD}], and in other ways.

The lens space values given above can be combined to give those on
the other quotients by using the cyclic decomposition which reads
here
  $$
  \eta_{\Ga'}(s)
  ={1\over2}\bigg(\sum_q \eta_{Z_{2q}}(s)-\eta_{Z_2}(s)\bigg)\,.
  \eql{etadecomp}
  $$
The numbers evaluated using this relation provide a useful check.

It should be mentioned that Seade, [\pref{Seade}], has looked at the
$\eta$ invariant on the factored three--sphere and, more recently,
Cisneros--Molina, [\pref{CM}], has extended the discussion to the
twisted case. General calculations can be found in Goette,
[\pref{Goette1}].
\section{\bf 12. Discussion and conclusion.}

A number of points arising can be mentioned. Concerning the cosecant
sums, (\peq{cosecs}), the fact that they are rational numbers for the
cyclic and dihedral cases follows from a residue evaluation. For the
other groups it is not so evident directly, but follows from the
cyclic decomposition.

Our discussion of spinors was restricted to the natural, trivial spin
structure on the factored three--sphere. The dependence of the
spectrum on the spin structures is discussed in general by B\"ar,
[\pref{Bar1,Bar2}] who also considers the squashed (Berger) sphere.

It is also possible to calculate the functional determinants for
spinors on S$^3/\Ga'$ and this will be given at another time. The
full sphere results exist already. The evaluation for the Maxwell
field is complicated by the non--zero value of $\ze_M(0)$.

On homogeneous quotients of the Einstein Universe, the vacuum energy
density, $\av {T_0^0}$, is obtained simply by dividing $E$ by the
volume. It is also possible to obtain the spatial densities
$\av{T_i^j}$, [\pref{DandB}]. Because the symmetry group is generally
reduced, these contains geometric structure over and above that
arising from the metric.

The case of non--homogeneous quotients is much harder but in certain
circumstances an exact $\av {T_0^0}$ can be found with some work,
[\pref{Jadhav}].

As mentioned, it is possible to introduce an equivariant twisting
according to ${\rm Hom}\big(\Ga,U(N)\big)$, say. The analysis is
one in character theory. The scalar summations can still be performed
in the case of one--sided lens spaces and result in generalised Bernoulli
polynomials.

The construction of the eigenfunctions is left aside as a chapter in
the theory of symmetry adaptation most familiar, perhaps, in solid
state physics.

\newpage

\section{\bf References.}
\begin{putreferences}
  \ref{DandA}{Dowker,J.S. and Apps, J.S. \cqg{}{}{}.}
  \ref{Weil}{Weil,A., {\it Elliptic functions according to Eisenstein
  and Kronecker}, Springer, Berlin, 1976.}
  \ref{Ling}{Ling,C-H. {\it SIAM J.Math.Anal.} {\bf5} (1974) 551.}
  \ref{Ling2}{Ling,C-H. {\it J.Math.Anal.Appl.}(1988).}
 \ref{BMO}{Brevik,I., Milton,K.A. and Odintsov, S.D. {\it Entropy bounds in
 $R\times S^3$ geometries}. hep-th/0202048.}
 \ref{KandL}{Kutasov,D. and Larsen,F. {\it JHEP} 0101 (2001) 1.}
 \ref{KPS}{Klemm,D., Petkou,A.C. and Siopsis {\it Entropy
 bounds, monoticity properties and scaling in CFT's}. hep-th/0101076.}
 \ref{DandC}{Dowker,J.S. and Critchley,R. \prD{15}{1976}{1484}.}
 \ref{AandD}{Al'taie, M.B. and Dowker, J.S. \prD{18}{1978}{3557}.}
 \ref{Dow1}{Dowker,J.S. \prD{37}{1988}{558}.}
 \ref{Dow3}{Dowker,J.S. \prD{28}{1983}{3013}.}
 \ref{DandK}{Dowker,J.S. and Kennedy,G. \jpa{}{1978}{}.}
 \ref{Dow2}{Dowker,J.S. \cqg{1}{1984}{359}.}
 \ref{DandKi}{Dowker,J.S. and Kirsten, K.{\it Comm. in Anal. and Geom.
 }{\bf7}(1999) 641.}
 \ref{DandKe}{Dowker,J.S. and Kennedy,G.\jpa{11}{1978}{895}.}
 \ref{Gibbons}{Gibbons,G.W. \pl{60A}{1977}{385}.}
 \ref{Cardy}{Cardy,J.L. \np{366}{1991}{403}.}
 \ref{ChandD}{Chang,P. and Dowker,J.S. \np{395}{1993}{407}.}
 \ref{DandC2}{Dowker,J.S. and Critchley,R. \prD{13}{1976}{224}.}
 \ref{Camporesi}{Camporesi,R. \prp{196}{1990}{1}.}
 \ref{BandM}{Brown,L.S. and Maclay,G.J. \pr{184}{1969}{1272}.}
 \ref{CandD}{Candelas,P. and Dowker,J.S. \prD{19}{1979}{2902}.}
 \ref{Unwin1}{Unwin,S.D. Thesis. University of Manchester. 1979.}
 \ref{Unwin2}{Unwin,S.D. \jpa{13}{1980}{313}.}
 \ref{DandB}{Dowker,J.S.and Banach,R. \jpa{11}{1978}{2255}.}
 \ref{Obhukov}{Obhukov,Yu.N. \pl{109B}{1982}{195}.}
 \ref{Kennedy}{Kennedy,G. \prD{23}{1981}{2884}.}
 \ref{CandT}{Copeland,E. and Toms,D.J. \np {255}{1985}{201}.}
 \ref{ELV}{Elizalde,E., Lygren, M. and Vassilevich,
 D.V. \jmp {37}{1996}{3105}.}
 \ref{Malurkar}{Malurkar,S.L. {\it J.Ind.Math.Soc} {\bf16} (1925/26) 130.}
 \ref{Glaisher}{Glaisher,J.W.L. {\it Messenger of Math.} {\bf18}
(1889) 1.} \ref{Anderson}{Anderson,A. \prD{37}{1988}{536}.}
 \ref{CandA}{Cappelli,A. and D'Appollonio,\pl{487B}{2000}{87}.}
 \ref{Wot}{Wotzasek,C. \jpa{23}{1990}{1627}.}
 \ref{RandT}{Ravndal,F. and Tollesen,D. \prD{40}{1989}{4191}.}
 \ref{SandT}{Santos,F.C. and Tort,A.C. \pl{482B}{2000}{323}.}
 \ref{FandO}{Fukushima,K. and Ohta,K. {\it Physica} {\bf A299} (2001) 455.}
 \ref{GandP}{Gibbons,G.W. and Perry,M. \prs{358}{1978}{467}.}
 \ref{Dow4}{Dowker,J.S. {\it Zero modes, entropy bounds and partition
functions.} hep-th\break /0203026.}
  \ref{Rad}{Rademacher,H. {\it Topics in analytic number theory,}
Springer-Verlag,  Berlin,1973.}
  \ref{Halphen}{Halphen,G.-H. {\it Trait\'e des Fonctions Elliptiques}, Vol 1,
Gauthier-Villars, Paris, 1886.}
  \ref{CandW}{Cahn,R.S. and Wolf,J.A. {\it Comm.Mat.Helv.} {\bf 51} (1976) 1.}
  \ref{Berndt}{Berndt,B.C. \rmjm{7}{1977}{147}.}
  \ref{Hurwitz}{Hurwitz,A. \ma{18}{1881}{528}.}
  \ref{Hurwitz2}{Hurwitz,A. {\it Mathematische Werke} Vol.I. Basel,
  Birkhauser, 1932.}
  \ref{Berndt2}{Berndt,B.C. \jram{303/304}{1978}{332}.}
  \ref{RandA}{Rao,M.B. and Ayyar,M.V. \jims{15}{1923/24}{150}.}
  \ref{Hardy}{Hardy,G.H. \jlms{3}{1928}{238}.}
  \ref{TandM}{Tannery,J. and Molk,J. {\it Fonctions Elliptiques},
   Gauthier-Villars, Paris, 1893--1902.}
  \ref{schwarz}{Schwarz,H.-A. {\it Formeln und Lehrs\"atzen zum Gebrauche..},
  Springer 1893.(The first edition was 1885.) The French translation by
Henri Pad\'e is {\it Formules et Propositions pour L'Emploi...},
Gauthier-Villars, Paris, 1894}
  \ref{Hancock}{Hancock,H. {\it Theory of elliptic functions}, Vol I.
   Wiley, New York 1910.}
  \ref{watson}{Watson,G.N. \jlms{3}{1928}{216}.}
  \ref{MandO}{Magnus,W. and Oberhettinger,F. {\it Formeln und S\"atze},
  Springer-Verlag, Berlin 1948.}
  \ref{Klein}{Klein,F. {\it Lectures on the Icosohedron}
  (Methuen, London, 1913).}
  \ref{AandL}{Appell,P. and Lacour,E. {\it Fonctions Elliptiques},
  Gauthier-Villars,
  Paris, 1897.}
  \ref{HandC}{Hurwitz,A. and Courant,C. {\it Allgemeine Funktionentheorie},
  Springer,
  Berlin, 1922.}
  \ref{WandW}{Whittaker,E.T. and Watson,G.N. {\it Modern analysis},
  Cambridge 1927.}
  \ref{SandC}{Selberg,A. and Chowla,S. \jram{227}{1967}{86}. }
  \ref{zucker}{Zucker,I.J. {\it Math.Proc.Camb.Phil.Soc} {\bf 82 }(1977) 111.}
  \ref{glasser}{Glasser,M.L. {\it Maths.of Comp.} {\bf 25} (1971) 533.}
  \ref{GandW}{Glasser, M.L. and Wood,V.E. {\it Maths of Comp.} {\bf 25} (1971)
  535.}
  \ref{greenhill}{Greenhill,A,G. {\it The Applications of Elliptic
  Functions}, MacMillan, London, 1892.}
  \ref{Weierstrass}{Weierstrass,K. {\it J.f.Mathematik (Crelle)}
{\bf 52} (1856) 346.}
  \ref{Weierstrass2}{Weierstrass,K. {\it Mathematische Werke} Vol.I,p.1,
  Mayer u. M\"uller, Berlin, 1894.}
  \ref{Fricke}{Fricke,R. {\it Die Elliptische Funktionen und Ihre Anwendungen},
    Teubner, Leipzig. 1915, 1922.}
  \ref{Konig}{K\"onigsberger,L. {\it Vorlesungen \"uber die Theorie der
 Elliptischen Funktionen},  \break Teubner, Leipzig, 1874.}
  \ref{Milne}{Milne,S.C. {\it The Ramanujan Journal} {\bf 6} (2002) 7-149.}
  \ref{Schlomilch}{Schl\"omilch,O. {\it Ber. Verh. K. Sachs. Gesell. Wiss.
  Leipzig}  {\bf 29} (1877) 101-105; {\it Compendium der h\"oheren Analysis},
  Bd.II, 3rd Edn, Vieweg, Brunswick, 1878.}
  \ref{BandB}{Briot,C. and Bouquet,C. {\it Th\`eorie des Fonctions
  Elliptiques}, Gauthier-Villars, Paris, 1875.}
  \ref{Dumont}{Dumont,D. \aim {41}{1981}{1}.}
  \ref{Andre}{Andr\'e,D. {\it Ann.\'Ecole Normale Superior} {\bf 6} (1877) 265;
  {\it J.Math.Pures et Appl.} {\bf 5} (1878) 31.}
  \ref{Raman}{Ramanujan,S. {\it Trans.Camb.Phil.Soc.} {\bf 22} (1916) 159;
 {\it Collected Papers}, Cambridge, 1927}
  \ref{Weber}{Weber,H.M. {\it Lehrbuch der Algebra} Bd.III, Vieweg,
  Brunswick 190  3.}
  \ref{Weber2}{Weber,H.M. {\it Elliptische Funktionen und algebraische Zahlen},
  Vieweg, Brunswick 1891.}
  \ref{ZandR}{Zucker,I.J. and Robertson,M.M.
  {\it Math.Proc.Camb.Phil.Soc} {\bf 95 }(1984) 5.}
  \ref{JandZ1}{Joyce,G.S. and Zucker,I.J.
  {\it Math.Proc.Camb.Phil.Soc} {\bf 109 }(1991) 257.}
  \ref{JandZ2}{Zucker,I.J. and Joyce.G.S.
  {\it Math.Proc.Camb.Phil.Soc} {\bf 131 }(2001) 309.}
  \ref{zucker2}{Zucker,I.J. {\it SIAM J.Math.Anal.} {\bf 10} (1979) 192,}
  \ref{BandZ}{Borwein,J.M. and Zucker,I.J. {\it IMA J.Math.Anal.} {\bf 12}
  (1992) 519.}
  \ref{Cox}{Cox,D.A. {\it Primes of the form $x^2+n\,y^2$}, Wiley, New York,
  1989.}
  \ref{BandCh}{Berndt,B.C. and Chan,H.H. {\it Mathematika} {\bf42} (1995) 278.}
  \ref{EandT}{Elizalde,R. and Tort.hep-th/}
  \ref{KandS}{Kiyek,K. and Schmidt,H. {\it Arch.Math.} {\bf 18} (1967) 438.}
  \ref{Oshima}{Oshima,K. \prD{46}{1992}{4765}.}
  \ref{greenhill2}{Greenhill,A.G. \plms{19} {1888} {301}.}
  \ref{Russell}{Russell,R. \plms{19} {1888} {91}.}
  \ref{BandB}{Borwein,J.M. and Borwein,P.B. {\it Pi and the AGM}, Wiley,
  New York, 1998.}
  \ref{Resnikoff}{Resnikoff,H.L. \tams{124}{1966}{334}.}
  \ref{vandp}{Van der Pol, B. {\it Indag.Math.} {\bf18} (1951) 261,272.}
  \ref{Rankin}{Rankin,R.A. {\it Modular forms} CUP}
  \ref{Rankin2}{Rankin,R.A. {\it Proc. Roy.Soc. Edin.} {\bf76 A} (1976) 107.}
  \ref{Skoruppa}{Skoruppa,N-P. {\it J.of Number Th.} {\bf43} (1993) 68 .}
  \ref{Down}{Dowker.J.S. \np {104}{2002}{153}.}
  \ref{Eichler}{Eichler,M. \mz {67}{1957}{267}.}
  \ref{Zagier}{Zagier,D. \invm{104}{1991}{449}.}
  \ref{Lang}{Lang,S. {\it Modular Forms}, Springer, Berlin, 1976.}
  \ref{Kosh}{Koshliakov,N.S. {\it Mess.of Math.} {\bf 58} (1928) 1.}
  \ref{BandH}{Bodendiek, R. and Halbritter,U. \amsh{38}{1972}{147}.}
  \ref{Smart}{Smart,L.R., \pgma{14}{1973}{1}.}
  \ref{Grosswald}{Grosswald,E. {\it Acta. Arith.} {\bf 21} (1972) 25.}
  \ref{Kata}{Katayama,K. {\it Acta Arith.} {\bf 22} (1973) 149.}
  \ref{Ogg}{Ogg,A. {\it Modular forms and Dirichlet series} (Benjamin,
  New York,
   1969).}
  \ref{Bol}{Bol,G. \amsh{16}{1949}{1}.}
  \ref{Epstein}{Epstein,P. \ma{56}{1903}{615}.}
  \ref{Petersson}{Petersson.}
  \ref{Serre}{Serre,J-P. {\it A Course in Arithmetic}, Springer,
  New York, 1973.}
  \ref{Schoenberg}{Schoenberg,B., {\it Elliptic Modular Functions},
  Springer, Berlin, 1974.}
  \ref{Apostol}{Apostol,T.M. \dmj {17}{1950}{147}.}
  \ref{Ogg2}{Ogg,A. {\it Lecture Notes in Math.} {\bf 320} (1973) 1.}
  \ref{Knopp}{Knopp,M.I. \dmj {45}{1978}{47}.}
  \ref{Knopp2}{Knopp,M.I. \invm {}{1994}{361}.}
  \ref{LandZ}{Lewis,J. and Zagier,D. \aom{153}{2001}{191}.}
  \ref{DandK1}{Dowker,J.S. and Kirsten,K. {\it Elliptic functions and
  temperature inversion symmetry on spheres} hep-th/.}
  \ref{HandK}{Husseini and Knopp.}
  \ref{Kober}{Kober,H. \mz{39}{1934-5}{609}.}
  \ref{HandL}{Hardy,G.H. and Littlewood, \am{41}{1917}{119}.}
  \ref{Watson}{Watson,G.N. \qjm{2}{1931}{300}.}
  \ref{SandC2}{Chowla,S. and Selberg,A. {\it Proc.Nat.Acad.} {\bf 35}
  (1949) 371.}
  \ref{Landau}{Landau, E. {\it Lehre von der Verteilung der Primzahlen},
  (Teubner, Leipzig, 1909).}
  \ref{Berndt4}{Berndt,B.C. \tams {146}{1969}{323}.}
  \ref{Berndt3}{Berndt,B.C. \tams {}{}{}.}
  \ref{Bochner}{Bochner,S. \aom{53}{1951}{332}.}
  \ref{Weil2}{Weil,A.\ma{168}{1967}{}.}
  \ref{CandN}{Chandrasekharan,K. and Narasimhan,R. \aom{74}{1961}{1}.}
  \ref{Rankin3}{Rankin,R.A. {} {} ().}
  \ref{Berndt6}{Berndt,B.C. {\it Trans.Edin.Math.Soc}.}
  \ref{Elizalde}{Elizalde,E. {\it Ten Physical Applications of Spectral
  Zeta Function Theory}, \break (Springer, Berlin, 1995).}
  \ref{Allen}{Allen,B., Folacci,A. and Gibbons,G.W. \pl{189}{1987}{304}.}
  \ref{Krazer}{Krazer}
  \ref{Elizalde3}{Elizalde,E. {\it J.Comp.and Appl. Math.} {\bf 118}
  (2000) 125.}
  \ref{Elizalde2}{Elizalde,E., Odintsov.S.D, Romeo, A. and Bytsenko,
  A.A and
  Zerbini,S.
  {\it Zeta function regularisation}, (World Scientific, Singapore,
  1994).}
  \ref{Eisenstein}{Eisenstein}
  \ref{Hecke}{Hecke,E. \ma{112}{1936}{664}.}
  \ref{Terras}{Terras,A. {\it Harmonic analysis on Symmetric Spaces} (Springer,
  New York, 1985).}
  \ref{BandG}{Bateman,P.T. and Grosswald,E. {\it Acta Arith.} {\bf 9}
  (1964) 365.}
  \ref{Deuring}{Deuring,M. \aom{38}{1937}{585}.}
  \ref{Guinand}{Guinand.}
  \ref{Guinand2}{Guinand.}
  \ref{Minak}{Minakshisundaram.}
  \ref{Mordell}{Mordell,J. \prs{}{}{}.}
  \ref{GandZ}{Glasser,M.L. and Zucker, {}.}
  \ref{Landau2}{Landau,E. \jram{}{1903}{64}.}
  \ref{Kirsten1}{Kirsten,K. \jmp{35}{1994}{459}.}
  \ref{Sommer}{Sommer,J. {\it Vorlesungen \"uber Zahlentheorie}
  (1907,Teubner,Leipzig).
  French edition 1913 .}
  \ref{Reid}{Reid,L.W. {\it Theory of Algebraic Numbers},
  (1910,MacMillan,New York).}
  \ref{Milnor}{Milnor, J. {\it Is the Universe simply--connected?},
  IAS, Princeton, 1978.}
  \ref{Milnor2}{Milnor, J. \ajm{79}{1957}{623}.}
  \ref{Opechowski}{Opechowski,W. {\it Physica} {\bf 7} (1940) 552.}
  \ref{Bethe}{Bethe, H.A. \zfp{3}{1929}{133}.}
  \ref{LandL}{Landau, L.D. and Lishitz, E.M. {\it Quantum
  Mechanics} (Pergamon Press, London, 1958).}
  \ref{GPR}{Gibbons, G.W., Pope, C. and R\"omer, H., \np{B157}{1979}{377}.}
  \ref{Jadhav}{Jadhav, S. PhD Thesis, University of Manchester 1990.}
  \ref{DandJ}{Dowker,J.S. and Jadhav, S. \prD{39}{1989}{1196}.}
  \ref{CandM}{Coxeter, H.S.M. and Moser, W.O.J. {\it Generators and
  relations of finite groups} Springer. Berlin. 1957.}
  \ref{Coxeter2}{Coxeter, H.S.M. {\it Regular Complex Polytopes},
   (Cambridge University Press,
  Cambridge, 1975).}
  \ref{Coxeter}{Coxeter, H.S.M. {\it Regular Polytopes}.}
  \ref{Stiefel}{Stiefel, E., J.Research NBS {\bf 48} (1952) 424.}
  \ref{BandS}{Brink and Satchler {\it Angular momentum theory}.}
  \ref{Rose}{Rose}
  \ref{Schwinger}{Schwinger,J.}
  \ref{Bromwich}{Bromwich, T.J.I'A. {\it Infinite Series},
  (Macmillan, 1947).}
  \ref{Ray}{Ray,D.B. \aim{4}{1970}{109}.}
  \ref{Ikeda}{Ikeda,A. {\it Kodai Math.J.} {\bf 18} (1995) 57.}
  \ref{Kennedy}{Kennedy,G. \prD{23}{1981}{2884}.}
  \ref{Ellis}{Ellis,G.F.R. {\it General Relativity} {\bf2} (1971) 7.}
  \ref{Dow8}{Dowker,J.S. \cqg{20}{2003}{L105}.}
  \ref{IandY}{Ikeda, A and Yamamoto, Y. \ojm {16}{1979}{447}.}
  \ref{BandI}{Bander,M. and Itzykson,C. \rmp{18}{1966}{2}.}
  \ref{Schulman}{Schulman, L.S. \prD{176}{1968}{1558}.}
  \ref{Bar1}{B\"ar,C. {\it Arch.d.Math.}{\bf 59} (1992) 65.}
  \ref{Bar2}{B\"ar,C. {\it Geom. and Func. Anal.} {\bf 6} (1996) 899.}
  \ref{Vilenkin}{Vilenkin, N.J. {\it Special functions},
  (Am.Math.Soc., Providence, 1968).}
  \ref{Talman}{Talman, J.D. {\it Special functions} (Benjamin,N.Y.,1968).}
  \ref{Miller}{Miller,W. {\it Symmetry groups and their applications}
  (Wiley, N.Y., 1972).}
  \ref{Dow3}{Dowker,J.S. \cmp{162}{94}{633}.}
  \ref{Cheeger}{Cheeger, J. \jdg {18}{83}{575}.}
  \ref{Dow6}{Dowker,J.S. \jmp{30}{1989}{770}.}
  \ref{Dow9}{Dowker,J.S. \jmp{42}{2001}{1501}.}
  \ref{Dow7}{Dowker,J.S. \jpa{25}{1992}{2641}.}
  \ref{Warner}{Warner.N.P. \prs{383}{1982}{379}.}
  \ref{Wolf}{Wolf, J.A. {\it Spaces of constant curvature},
  (McGraw--Hill,N.Y., 1967).}
  \ref{Meyer}{Meyer,B. \cjm{6}{1954}{135}.}
  \ref{BandB}{B\'erard,P. and Besson,G. {\it Ann. Inst. Four.} {\bf 30}
  (1980) 237.}
  \ref{PandM}{Polya,G. and Meyer,B. \cras{228}{1948}{28}.}
  \ref{Springer}{Springer, T.A. Lecture Notes in Math. vol 585 (Springer,
  Berlin,1977).}
  \ref{SeandT}{Threlfall, H. and Seifert, W. \ma{104}{1930}{1}.}
  \ref{Hopf}{Hopf,H. \ma{95}{1925}{313}. }
  \ref{Dow}{Dowker,J.S. \jpa{5}{1972}{936}.}
  \ref{LLL}{Lehoucq,R., Lachi\'eze-Rey,M. and Luminet, J.--P. {\it
  Astron.Astrophys.} {\bf 313} (1996) 339.}
  \ref{LaandL}{Lachi\'eze-Rey,M. and Luminet, J.--P.
  \prp{254}{1995}{135}.}
  \ref{Schwarzschild}{Schwarzschild, K., {\it Vierteljahrschrift der
  Ast.Ges.} {\bf 35} (1900) 337.}
  \ref{Starkman}{Starkman,G.D. \cqg{15}{1998}{2529}.}
  \ref{LWUGL}{Lehoucq,R., Weeks,J.R., Uzan,J.P., Gausman, E. and
  Luminet, J.--P. \cqg{19}{2002}{4683}.}
  \ref{Dow10}{Dowker,J.S. \prD{28}{1983}{3013}.}
  \ref{BandD}{Banach, R. and Dowker, J.S. \jpa{12}{1979}{2527}.}
  \ref{Jadhav2}{Jadhav,S. \prD{43}{1991}{2656}.}
  \ref{Gilkey}{Gilkey,P.B. {\it Invariance theory,the heat equation and
  the Atiyah--Singer Index theorem} (CRC Press, Boca Raton, 1994).}
  \ref{BandY}{Berndt,B.C. and Yeap,B.P. {\it Adv. Appl. Math.}
  {\bf29} (2002) 358.}
  \ref{HandR}{Hanson,A.J and R\"omer,H. \pl{80B}{1978}{58}.}
  \ref{Hill}{Hill,M.J.M. {\it Trans.Camb.Phil.Soc.} {\bf 13} (1883) 36.}
  \ref{Cayley}{Cayley,A. {\it Quart.Math.J.} {\bf 7} (1866) 304.}
  \ref{Seade}{Seade,J.A. {\it Anal.Inst.Mat.Univ.Nac.Aut\'on
  M\'exico} {\bf 21} (1981) 129.}
  \ref{CM}{Cisneros--Molina,J.L. {\it Geom.Dedicata} {\bf84} (2001)
  \ref{Goette1}{Goette,S. \jram {526} {2000} 181.}
  207.}
  \ref{NandO}{Nash,C. and O'Connor,D--J,, \jmp {36}{1995}{1462}.}
\end{putreferences}

\bye